\documentclass[submission, Phys]{SciPost}

\usepackage{caption}
\usepackage{subcaption}

\usepackage[utf8]{inputenc} % input encodings (allow UTF-8 input)
\usepackage[T1]{fontenc} 	% selecting font encodings (use 8-bit T1 fonts)
\usepackage[english]{babel} % multilingual support (English language/hyphenation)

\usepackage[bitstream-charter]{mathdesign}
\usepackage{geometry} 		% flexible and complete interface to document dimensions
\usepackage{amsmath} 		% math package (American Mathematical Society)
\usepackage{mathtools} 		% math package (fixes various deficiencies of amsmath)
\usepackage{easyReview}
\usepackage{float} 			% floating objects such as figures and tables
\usepackage{graphicx} 		% enhanced support for graphics
\usepackage{tabularx} 		% tabulars with adjustable-width columns
\usepackage{booktabs} 		% professional-quality tables
\usepackage{color, xcolor} 	% foreground and background colour management
\usepackage{pdfpages} 		% inclusion of external multi-page PDF documents
\usepackage{extarrows} 		% extra arrows beyond those provided in amsmath
\usepackage{multirow} 		% create tabular cells spanning multiple rows
\usepackage{multicol} 		% intermix single and multiple columns
\usepackage{enumitem} 		% control layout of itemize, enumerate, description
\usepackage{xspace} 		% define commands that appear not to eat spaces
\usepackage{stackrel} 		% enhancement to the \stackrel command
\usepackage{tikz} 			% create PostScript and PDF graphics
\usepackage{braket} 		% Dirac bra-ket notation
\usepackage{bm} 			% access bold symbols in maths mode
\usepackage{tensor} 		% typeset tensors (tensor-style super- and subscripts)
\usepackage{slashed} 		% slash through characters (Feynman slash notation)
\usepackage{siunitx} 		% SI units package (typesetting values with units)
\usepackage{lastpage} 		% reference last page
\usepackage{cite} 			% improved citation handling
\usepackage[normalem]{ulem} % package for underlining
\usepackage{fontawesome} 	% access to web-related icons
\usepackage{tocloft} 		% control over the typography of the TOC, LOF, and LOT
\usepackage{titlesec} 		% interface to sectioning commands (various title styles)
\usepackage{doi} 			% create correct hyperlinks for DOI numbers
\usepackage{hyperref} 		% hypertext links (handel cross-referencing commands)
\usepackage[most]{tcolorbox} 					% coloured and framed text boxes
\usepackage[nameinlink, capitalize]{cleveref} 	% intelligent cross-referencing
\usepackage[nottoc, notlot, notlof]{tocbibind} 	% add references/index/contents to TOC
\usepackage[ruled, vlined]{algorithm2e} 		% floating algorithm environment
\usepackage{makecell}
\usepackage[makeroom]{cancel}
\usepackage{feynmf}

\makeatletter
\def\BState{\State\hskip-\ALG@thistlm}
\makeatother

\makeatletter
\@ifundefined{pdfoutput}{}{\DeclareGraphicsRule{*}{mps}{*}{}}
\makeatother

\makeatletter
\DeclareRobustCommand*{\bfseries}{%
   \not@math@alphabet\bfseries\mathbf
   \fontseries\bfdefault\selectfont
   \boldmath
}
\makeatother

\hypersetup{
	pdftitle={Unfolding without Iterations, Adversaries, or Surrogates},
	pdfauthor={Ayodele Ore and Tilman Plehn},
	colorlinks=true, 			% false: boxed links, true: colored links
	linkcolor={red!50!black}, 	% color of internal links (sections, pages, etc.)
	citecolor={blue!50!black}, 	% color of citation links (links to bibliography)
	urlcolor={blue!80!black} 	% color of URL links (external links)
} 
\DeclareSymbolFont{usualmathcal}{OMS}{cmsy}{m}{n}
\DeclareSymbolFontAlphabet{\mathcal}{usualmathcal}

\SetArgSty{textnormal}
\SetKwComment{Comment}{{\small\#}~}{}
\SetCommentSty{mycommfont}

\setitemize{itemsep=0pt, parsep=0pt} 				% adjust itemize environment
\setenumerate{itemsep=0pt, parsep=0pt} 				% adjust enumerate environment
\setitemize{itemsep=2pt,topsep=2pt,parsep=0pt,partopsep=0pt,leftmargin=*}
\setenumerate{itemsep=0pt,topsep=2pt,parsep=0pt,partopsep=0pt,labelindent=3pt,leftmargin=*}
\usepackage{amsmath}
 
\usepackage{amsthm} 		% math package (typesetting theorems using AMS style)
\theoremstyle{definition}

\usetikzlibrary{arrows, shapes.geometric, arrows.meta, shapes, decorations.pathreplacing, fit}

\definecolor{Rcolor}{HTML}{E99595}
\definecolor{Gcolor}{HTML}{C5E0B4}
\definecolor{Bcolor}{HTML}{9DC3E6}
\definecolor{Ycolor}{HTML}{FFE699}

\tikzstyle{expr} = [rectangle, rounded corners=0.3ex, minimum width=1.8cm, minimum height=1.5cm, text centered, align=center, inner sep=0, fill=white, font=\LARGE, draw]
\tikzstyle{txt_huge} = [align=center, font=\Huge, scale=2]
\tikzstyle{txt} = [align=center, font=\LARGE]
\tikzstyle{cinn} = [double arrow, double arrow head extend=0cm, double arrow tip angle=130, shape border rotate=90, inner sep=0, align=center, minimum width=2.3cm, minimum height=2.5cm, fill=Gcolor, draw, font=\LARGE]
\tikzstyle{cinn_black} = [cinn, minimum height=2.7cm, fill=black]
\tikzstyle{arrow} = [thick,-{Latex[scale=1.0]}, line width=0.2mm, color=black]
\tikzstyle{line} = [thick, line width=0.2mm, color=black]
\definecolor{red_cb}{HTML}{e41a1c}
\definecolor{blue_cb}{HTML}{377eb8}
\definecolor{green_cb}{HTML}{4daf4a}
\definecolor{purple_cb}{HTML}{984ea3}
\definecolor{orange_cb}{HTML}{ff7f00}

\definecolor{EmeraldGreen}{HTML}{1ea78d}
\definecolor{EnglishRed}{HTML}{b02427}
\hypersetup{colorlinks=true,urlcolor=EmeraldGreen,citecolor=EmeraldGreen,linkcolor=EnglishRed}

\newcommand{\ie}{\text{i.e.}\;}

\newcommand{\pd}{p_\text{data}}
\newcommand{\ps}{p_\text{sim}}
\newcommand{\pu}{p_\text{unfold}}

\newcommand{\Rbar}{\ensuremath{\overline{R}}\xspace}

\newcommand{\XXLangle}{\biggl\langle}
\newcommand{\XXRangle}{\biggr\rangle}

\newcommand{\qqquad}{\qquad\quad}

\newcommand\one{\leavevmode\hbox{\small1\normalsize\kern-.33em1}}
\newcommand{\loss}{\mathcal{L}} 	% loss value

\newcommand{\arXiv}[2][]{%
	\ifthenelse{\equal{#1}{}}%
	{\href{http://arxiv.org/abs/#2}{arXiv:#2}}%
	{\href{http://arxiv.org/abs/#2}{arXiv:#2~[#1]}}}

\def\slashchar#1{\setbox0=\hbox{$#1$}           % set a box for #1
   \dimen0=\wd0                                 % and get its size
   \setbox1=\hbox{/} \dimen1=\wd1               % get size of /
   \ifdim\dimen0>\dimen1                        % #1 is bigger
      \rlap{\hbox to \dimen0{\hfil/\hfil}}      % so center / in box
      #1                                        % and print #1
   \else                                        % / is bigger
      \rlap{\hbox to \dimen1{\hfil$#1$\hfil}}   % so center #1
      /                                         % and print /
   \fi}

\newcommand{\tikznode}[2]{%
\ifmmode%
\tikz[remember picture,baseline=(#1.base),inner sep=0pt] \node (#1) {$#2$};%
\else
\tikz[remember picture,baseline=(#1.base),inner sep=0pt] \node (#1) {#2};%
\fi}

\def\mathswitchr#1{\relax\ifmmode{\mathrm{#1}}\else$\mathrm{#1}$\xspace\fi}
\def\mathswitch#1{\relax\ifmmode#1\else$#1$\xspace\fi}

\usepackage[most]{tcolorbox}
\usepackage{listings}
\usepackage[table]{xcolor}

\definecolor{docnavy}{HTML}{002B49}   % #002B49
\definecolor{navykeyword}{HTML}{0A3A5A}
\definecolor{navycomment}{HTML}{5C6F7B}
\definecolor{navystring}{HTML}{7A5C5C}
\definecolor{navybg}{HTML}{F4F7FA}
\definecolor{navyframe}{HTML}{C9D6E2}
\definecolor{heatgreen}{HTML}{2E8B57}

\lstdefinestyle{navyblackpython}{
    language=Python,
    backgroundcolor=\color{navybg},
    basicstyle=\ttfamily\fontsize{8pt}{10pt}\selectfont\color{black}, % ← plain tokens in black
    keywordstyle=\color{docnavy}\bfseries,
    commentstyle=\color{navycomment}\itshape,
    stringstyle=\color{navystring},
    identifierstyle=\color{black},
    breaklines=true,
    showstringspaces=false,
    tabsize=4,
    frame=none,
}

\newtcblisting{pycode}[2][]{
    listing engine=listings,
    enhanced,
    listing only,
    title={\strut #2},
    listing options={style=navyblackpython},
    colframe=black,
    colback=navybg,
    coltitle=black,
    colbacktitle=navybg!80!black,
    fonttitle=\bfseries,
    boxrule=0.6pt,
    arc=6pt,
    top=10pt,
    bottom=8pt,
    left=8pt,
    right=8pt,
    toptitle=4pt,
    bottomtitle=2pt,
    before skip=8pt,
    after skip=8pt,
    width=0.9\linewidth,
    halign=center,
    #1
}
\graphicspath{{./figures/}}

\begin{document}

% article title
\begin{center}
    {\Large\textbf{Unfolding without Iterations, Adversaries, or Surrogates}}
\end{center}

% write the author list here, use first name (+ other initials) + surname
% format, separate subsequent authors by a comma, omit comma and use "and"
% for the last author (mark the corresponding author with a superscript star)
\begin{center}
    Ayodele Ore\textsuperscript{1} and Tilman Plehn\textsuperscript{1,2}
\end{center}

\begin{center}
    {\bf 1} Institut f\"ur Theoretische Physik, Universit\"at
    Heidelberg, Germany\\
    {\bf 2} Interdisciplinary Center for Scientific Computing (IWR),
    Universit\"at Heidelberg,
    Germany
\end{center}

\begin{center}
    \today
\end{center}

\section*{Abstract}
{\bf Correcting measurements for detector effects and constructing
    appropriate public data representations is a pressing problem in
    LHC physics. Current methods solve this inverse problem by relying on
    iterations, minimax optimization, or a surrogate forward mapping.
    We introduce Adversary-free Unfolding SanS Iteration or Emulation
    (AUSSIE), which
    dispenses with these mechanisms while remaining asymptotically correct.
    AUSSIE replaces the second OmniFold step with a new loss function
    that directly yields solutions with minimal dependence on the
    reference simulation. We showcase AUSSIE on various unfolding tasks,
including full-phase-space jet substructure.}

% include a table of contents (optional)
\vspace{10pt}
\noindent\rule{\linewidth}{1pt}
\tableofcontents\thispagestyle{fancy}
\noindent\rule{\linewidth}{1pt}
\clearpage

%%%%%%%%%%%%%%%%%%%%%%%%%%%%%%%%%%%%%%%%%%%%%%%%%%%%%%%%%
\section{Introduction}
\label{sec:intro}

Percent-level measurements at the LHC have changed the way we think
about hadron collider physics. A central pillar of this precision
program is high-fidelity simulation, connecting Lagriangians with
parton-level theory predictions and
observables~\cite{Campbell:2022qmc}. However, the expense and
inaccessibility of these simulations is rapidly becoming a critical
bottleneck  --- it limits the number of
hypotheses that can be tested with simulation-based inference and
precludes the analysis of data outside each experimental
collaboration.

Unfolding provides an alternative strategy by defining appropriate
data representations for analyses. It enables efficient comparison to
a wide range of models across experiments and shifts the
computational cost from repeated detector simulation to a one-time
inference~\cite{Cowan:2002in}. Traditional unfolding methods,
however, rely on binning and do not scale to high-dimensional phase
spaces~\cite{DAgostini:2010hil,Schmitt:2012kp}.

Modern machine learning~\cite{Plehn:2022ftl} enables unbinned
unfolding in many dimensions~\cite{Huetsch:2024quz,Canelli:2025ybb}. Two broad
classes of approaches have emerged. Discriminative methods, like
OmniFold~\cite{Andreassen:2019cjw}, train classifiers to reweight a
simulated reference to the unfolded
distribution~\cite{Chan:2023tbf,Desai:2024yft,Zhu:2024drd,Falcao:2025jom}.
This technique has been deployed in several experimental
analyses~\cite{LHCb:2022rky,H1:2023fzk,ATLAS:2024rpl,ATLAS:2025qtv,CMS:2025sws,Electron-PositronAlliance:2025hze}.
Alternatively, generative neural networks can be trained to sample
unfolded
events~\cite{Bellagente:2019uyp,Bellagente:2020piv,Vandegar:2020yvw,Howard:2021pos,Shmakov:2023kjj,Diefenbacher:2023wec,Shmakov:2024gkd,Pazos:2024nfe,Butter:2024vbx,Favaro:2025psi,Butter:2025via,Petitjean:2025tgk},
with complementary advantages in terms of efficiency and flexibility.

A common part of many unfolding methods, both discriminative and generative, is the use of iterative
refinement to reduce bias toward the simulated prior and to ensure
closure at reconstruction
level~\cite{Andreassen:2019cjw,Backes:2022sph}. While iterations
guarantee convergence in principle, the rate of convergence may
be slow for substantial forward smearing. This
makes iterative strategies computationally expensive since they are
fundamentally sequential and cannot be parallelized. Furthermore, because
the number of required iterations is not known \emph{a priori}, stopping
criteria must be selected heuristically. As a result, residual bias
from the simulation can persist if the procedure is terminated
prematurely, while excessive iteration may lead to increased variance
or overfitting.

A few non-iterative approaches have also been explored. Empirical
Bayes approaches with learned forward
surrogates~\cite{Vandegar:2020yvw,Butter:2025via} avoid explicit
simulation dependence, but require expensive sampling and integration over the
latent phase space at each training step. Adversarial
formulations~\cite{Datta:2018mwd, Bellagente:2019uyp,Desai:2024yft}
train a generator on the latent phase space to fool a
reconstruction-level discriminator. While efficient, the non-convex
minimax optimization inherent to these methods tends
to be unstable and sensitive to hyperparameter choices.

We devise a novel approach to unfolding that is non-iterative,
accurate, and fast:   Adversary-free Unfolding SanS Iteration or
Emulation (AUSSIE). AUSSIE is a discriminative unfolding method
similar to OmniFold, but with a modified second step that
eliminates the need for iterative refinement. By directly finding
solutions to the unfolding problem, \ie latent distributions whose
forward mapping matches the data, AUSSIE exhibits drastically reduced
dependence on the reference simulation. We demonstrate that AUSSIE
consistently finds better solutions than ten OmniFold iterations, as
judged by the closure onto pseudodata in the reconstructed and latent
phase spaces.

The remainder of this paper is organized as follows. In
Section~\ref{sec:aussie}, we introduce the preliminaries of unfolding
and derive the AUSSIE training objective. Section~\ref{sec:res}
presents a series of results, starting with a toy Gaussian example
before progressing to jet substructure and full parton-level events.
In each case, we compare to OmniFold as a benchmark. Finally,
Section~\ref{sec:outlook} contains our concluding remarks. We also
include three appendices containing additional
technical details to facilitate reproducibility.

%%%%%%%%%%%%%%%%%%%%%%%%%%%%%%%%%%%%%%%%%%%%%%%%%%%%%%%%%
\section{Adversary-free Unfolding SanS Iteration or Emulation (AUSSIE)}
\label{sec:aussie}

Unfolding as an inverse problem deconvolves observations $x$
that arise from distortions of some underlying truth $z$, for
instance, through hadronization or detector interactions. For a
stochastic distortion, it involves the following probability densities:
\begin{alignat}{9}
    & \ps(z)
    \quad \xleftrightarrow{\text{unfolding inference}} \quad
    && \pu(z)
    \notag \\
    & \hspace*{-3mm} {\scriptstyle \ps(x|z)} \Bigg\downarrow
    && \hspace*{+6mm}
    \Bigg\uparrow {\scriptstyle\text{unfold}}
    \notag \\
    & \ps(x)
    \quad \xleftrightarrow{\text{\; forward inference \;}} \quad
    && \pd(x)
    \label{eq:schematic}
\end{alignat}
The latent variable $z$ may be at particle level or parton level, and
we use the term part level to cover both cases. The conditional
probability $\ps(x|z)$ is the transfer function
characterizing forward simulation applied to $z$.
On the simulation side this implies a forward folding,
\begin{align}
    \ps(x) = \int dz \; \ps(x|z) \; \ps(z) \; .
    \label{eq:integral-map}
\end{align}
In simulation, we have complete access to $z$ and can sample pairs
that implicitly encode the forward conditional distribution,
\begin{align}
    (z,x) \sim \ps(x,z) = \ps(x|z) \; \ps(z)\; .
\end{align}
On the data side we just have access to samples from $\pd(x)$. The
goal is to find a distribution $\pu(z)$ that maps forward to $\pd(x)$
under Eq.\,\eqref{eq:integral-map},
\begin{align}
    \pd(x) = \int dz \; \ps(x|z) \; \pu(z) \; .
    \label{eq:unfold-solution}
\end{align}
Assuming that the transfer function is an accurate model of the real
detector, a solution $\pu(z)$ can be interpreted as an
estimate for the part-level phase space distribution in data.
However, it should be understood that the forward process of the
detector acts on the full part-level phase space. As such, the
assumption may be violated even for perfect detector simulation if
effects that influence the forward process are not included
in~$z$~\cite{ATLAS:2024xxl,Butter:2025via,Butter:2025mek}. In a
similar vein, the definition of $\pu(z)$ is influenced by nuisance
parameters of the detector simulation~\cite{Chan:2023tbf,
Zhu:2024drd, Zhu:2025uyn,Valsecchi:2026kpp}. If the simulator is
differentiable, then
$\pu(z)$ can be obtained by maximizing the likelihood of observed
data with Eq.\,\eqref{eq:unfold-solution} by gradient descent.

%%%%%%%%%%%%%%%%%%%%%%%%%%%%%%%%%%%%%%%%%%%%%%%%%%%%%%%%%
\subsubsection*{Density ratio unfolding}

We propose Adversary-free Unfolding SanS Iteration or Emulation
(AUSSIE). Similar to OmniFold, AUSSIE phrases unfolding in terms of
density ratios. At reco level $x$ we define
\begin{align}
    R(x) = \frac{\pd(x)}{\ps(x)}  \; ,
    \label{eq:defr}
\end{align}
and the unfolding target condition in Eq.\,\eqref{eq:unfold-solution} becomes
\begin{align}
    R(x)
    = \frac{1}{\ps(x)}&\int dz \; \ps(x|z) \; \pu(z)
    \notag \\
    = &\int dz \; \ps(z|x) \; \frac{\pu(z)}{\ps(z)} \; .
\end{align}
If we define the corresponding part-level ratio as
\begin{align}
    \Rbar(z) \equiv \frac{\pu(z)}{\ps(z)}\;,
\end{align}
the above target relation in terms of density ratios reads
\begin{align}
    R(x) = \int dz \; \ps(z|x) \; \Rbar(z) \; .
    \label{eq:weight-map}
\end{align}
In this relation the conditional probability from the forward folding
in Eq.\,\eqref{eq:integral-map} is inverted. When unfolding with
density ratios, we use the estimate of $\Rbar(z)$ to construct the
unfolded distribution through reweighting
\begin{align}
    \pu(z) = \Rbar(z) \ps(z)\;.
    \label{eq:def_unfold}
\end{align}
%

%%%%%%%%%%%%%%%%%%%%%%%%%%%%%%%%%%%%%%%%%%%%%%%%%%%%%%%%%
\subsubsection*{Closure}

Because of the non-injective nature of the simulation and a potential
loss of information in the forward direction,
Eqs.\,\eqref{eq:unfold-solution} and~\eqref{eq:weight-map} admit
multiple solutions. Consequently, judging $\pu(z)$ in terms of
agreement with the part-level distribution in a pseudodata sample
can be misleading. Moreover, such a test cannot be performed in real
data. The part-level problem of ensuring complete coverage of the
solution space is an ongoing research question and beyond the scope
of our study.

Instead, we first assess the unfolding at reco-level, where a
two-sample test between $\pd(x)$ and the forward folding of $\pu(z)$ can
establish the validity of a solution. In terms of density ratios, we
realize this forward folding with the joint probability of part/reco pairs,
\begin{align}
    \int dz\, \ps(x,z)\,\Rbar(z)
    \overset{\text{Eq.\,\eqref{eq:weight-map}}}{=} R(x)\ps(x) = \pd(x)\;.
\end{align}
%

%%%%%%%%%%%%%%%%%%%%%%%%%%%%%%%%%%%%%%%%%%%%%%%%%%%%%%%%%
\subsubsection*{Training}

As a first step, AUSSIE approximates the reco-level density ratio by
training a classifier between samples from $\pd(x)$ and $\ps(x)$, for
example using the binary cross entropy (BCE) loss,
\begin{align}
    R_\theta(x) \approx R(x)\,.
    \label{eq:step-one-solution}
\end{align}
In the second step, AUSSIE trains a part-level network
$\Rbar_\varphi(z)$ such that it forward maps to $R_\theta(x)$ as
prescribed by Eq.\,\eqref{eq:weight-map}. To achieve this, we exploit a
maximum likelihood classifier (MLC)-like loss functional,
\begin{align}
    \loss_\text{MLC}[R_\theta, \Rbar_\varphi] &= \int dx dz \;
    \ps(x,z) \; \left[R_\theta(x) - \Rbar_\varphi(z)\log R_\theta(x)
    \right] \; ,
    \label{eq:mlc-loss}
\end{align}
which uses paired reco/part events $(x,z)$ from simulation. If we
freeze $\Rbar_\varphi(z)$, this is essentially a regression loss for
$R_\theta(x)$, as we can see from the functional derivative
\begin{align}
    G_{\theta,\varphi}(x) \equiv \frac{\delta
    \loss_\text{MLC}[R_\theta, \Rbar_\varphi]}{\delta R_\theta(x)}
    &= \ps(x) \left[1-\frac{1}{R_\theta(x)}\int dz \; \ps(z|x) \;
    \Rbar_\varphi(z)\right] \; .
    \label{eq:functional-derivative}
\end{align}
It vanishes everywhere exactly when the learned version of
Eq.\,\eqref{eq:weight-map} holds,
\begin{align}
    R_\theta(x) = \int dz \; \ps(z|x) \; \Rbar_\varphi(z)\;.
    \label{eq:weight-map-learned}
\end{align}
If $R_\theta(x)$ simultaneously satisfies
Eqs.\,\eqref{eq:weight-map-learned} and~\eqref{eq:step-one-solution},
then $\Rbar_\varphi(z)$ solves the unfolding problem.

Directly optimizing $\loss_\text{MLC}$ with respect to
$\Rbar_\varphi(z)$ does not yield the above unfolding relations, so we need
another way to use this result to train the unfolder. The key idea in
AUSSIE is to train $\Rbar_\varphi(z)$ such that the pretrained
classifier $R_\theta(x)$ lies at the stationary point of
$\loss_\text{MLC}$. To achieve this, we minimize a norm of
$G_{\theta,\varphi}$, now freezing $\theta$ and  varying $\varphi$.
For a function-space object, the natural choice of norm is the
Reproducing Kernel Hilbert Space (RKHS) norm
\begin{align}
    \lVert G_{\theta, \varphi}\rVert^2_\mathcal{H} \equiv \int dx dx'
    \; G_{\theta,\varphi}(x) K(x,x') G_{\theta,\varphi}(x')\;,
    \label{eq:rkhs_norm}
\end{align}
with $K$ a positive semi-definite kernel. $K$ should not be confused
with a detector response kernel, as it couples pairs of reco-level events.

The RKHS norm is minimized at zero when $G_{\theta, \varphi}(x)$ lies
in the null space of $K$. For a strictly positive-definite kernel
this occurs only when $G_{\theta, \varphi}(x)$ is pointwise zero,
implying that the unfolding condition is satisfied. Repeating the
above analysis with an appropriate BCE or Mean Squared Error (MSE)
loss recovers the same result.

AUSSIE performs unfolding by minimizing the RKHS norm, inverting the
integral relation from Eq.\,\eqref{eq:weight-map}. We consider the
following two options for implementing Eq.\,\eqref{eq:rkhs_norm}
as a loss function:
\begin{description}
    \item[Analytic kernel] We can use a closed-form and
        positive-definite kernel, for example a Gaussian
        \begin{align}
            K(x,x') = \exp \frac{-\lvert x-x'\rvert^2}{2\Lambda^2} \; ,
            \label{eq:kernel-gauss}
        \end{align}
        where $\Lambda$ is a scale hyperparameter. The RKHS-norm in
        Eq.\,\eqref{eq:rkhs_norm} can then be directly evaluated by
        rearranging Eq.\,\eqref{eq:functional-derivative} as a double
        expectation over paired part/reco samples
        \begin{align}
            \loss_\text{Gauss}[\Rbar_\varphi] = \left\langle
            \left(1-\frac{\Rbar_\varphi(z)}{R_\theta(x)}\right)
            K(x,x')
            \left(1-\frac{\Rbar_\varphi(z')}{R_\theta(x')}\right)\right\rangle_{\ps(x,z)\ps(x',z')}
            \label{eq:gauss-loss}
        \end{align}
        which follows by pulling out the integral and probability
        densities from $G_{\theta,\varphi}(x)$.

    \item[AutoDiff] Interestingly, taking the parameter gradient norm
        \begin{align}
            \big\lvert \nabla_\theta \loss_\text{MLC}\big\rvert^2
            =\sum_{\theta_i} \left( \nabla_{\theta_i}
            \loss_\text{MLC} [R_\theta, \Rbar_\varphi] \right)^2,
            \label{eq:autodiff-loss-l2}
        \end{align}
        is equivalent to Eq.\,\eqref{eq:rkhs_norm} with $K$ chosen as
        the Neural Tangent Kernel (NTK)~\cite{Jacot:2018ivh,Chiefa:2025cap},
        \begin{align}
            \big\lvert \nabla_\theta \loss_\text{MLC}\big\rvert^2
            &= \sum_{\theta_i} \left( \int dx \;
                \frac{\delta \loss_\text{MLC} [R_\theta,
                \Rbar_\varphi]}{\delta R_\theta(x)}\;\nabla_{\theta_i}
            R_\theta(x) \right)^2 \notag \\
            &= \sum_{\theta_i} \left( \int dx \; G_{\theta,\varphi}(x)\;
            \nabla_{\theta_i} R_\theta(x) \right)^2 \notag \\
            &= \int dx dx' \; G_{\theta,\varphi}(x)
            G_{\theta,\varphi}(x') \; \sum_{\theta_i}
            \nabla_{\theta_i} R_\theta(x) \nabla_{\theta_i} R_\theta(x')\;.
            \notag \\
            &= \lVert G_{\theta,\varphi} \rVert_\mathcal{H}^2
            \qqquad \text{with} \qqquad K(x, x')
            = \sum_{\theta_i} \nabla_{\theta_i} R_\theta(x) \;
            \nabla_{\theta_i} R_\theta(x') \; .
        \end{align}
        The NTK is strictly positive definite in the infinite-width
        limit but positive semi-definite for finite $\theta$.
        Therefore minimizing Eq.\,\eqref{eq:autodiff-loss-l2} only
        guarantees that $G_{\theta,\varphi}$ is pointwise zero if it
        is in the tangent space of $R_\theta(x)$, i.e. if the
        classifier is sufficiently expressive. We expect this to be
        satisfied since $G_{\theta,\varphi}$ is a simple function of
        $R_\theta(x)$ and we match the architecture with $\Rbar_\varphi(z)$.

        Numerically, we find that using the $L_1$ gradient norm
        \begin{align}
            \loss_\text{AutoDiff}[\Rbar_\varphi]
            &=\sum_{\theta_i} \left| \nabla_{\theta_i}
            \loss_\text{MLC} [R_\theta, \Rbar_\varphi] \right|
            \label{eq:autodiff-loss-l1}
        \end{align}
        gives better results, so we will use this loss function for
        our following studies. It shares the same optimal asymptotic limit
        as Eq.\,\eqref{eq:autodiff-loss-l2} despite not having a kernel
        interpretation. To evaluate $\loss_\text{AutoDiff}$,
        we apply the \texttt{autograd} engine in PyTorch
        to the Monte Carlo estimate of $\loss_\text{MLC}$. Further
        numerical details are provided in Appendix~\ref{app:numerics}.
\end{description}
These two options are complementary. The fixed analytic kernel is
computationally efficient, but introduces a hyperparameter and does
not translate easily to non-trivial representations of $x$, such as
point clouds. The AutoDiff loss is more expensive to compute, but has
no hyperparameters and exploits all symmetries encoded in the
architecture of $R_\theta(x)$. Given these considerations, we prefer
$\loss_\text{Gauss}$ for low-dimensional $x$ and
$\loss_\text{AutoDiff}$ for high-dimensional $x$.

%%%%%%%%%%%%%%%%%%%%%%%%%%%%%%%%%%%%%%%%%%%%%%%%%%%%%%%%%
\subsubsection*{OmniFold comparison}

It is instructive to briefly compare AUSSIE to OmniFold. OmniFold
also trains a classifier $\omega(x) \equiv R_\theta(x)$ to
learn the reco-level density ratio as in Eq.\,\eqref{eq:step-one-solution}.
In a second step, it trains $\nu(z) \equiv \Rbar_\varphi(z)$ to
regress $\omega(x)$, originally using a BCE loss. For consistency, we
write the equivalent MLC loss
\begin{align}
    \loss_\text{OmniFold}[\omega, \nu] &= \int dx dz \;
    \ps(x,z) \; \big[\nu(z) - \omega(x)\log
    \nu(z) \big] \; .
    \label{eq:omnifold-loss}
\end{align}
This leads to the following solution for OmniFold
\begin{align}
    \nu(z) = \int dx \, \ps(x|z)\; \omega(x)\;.
    \label{eq:app-omnifold-optimum}
\end{align}
This condition is an `inversion' of the unfolding target relation in
Eq.\,\eqref{eq:weight-map}. As a result, the forward folding of this
part-level ratio does not in general produce the correct reco-level ratio,
\begin{align}
    \omega(x) &\stackrel{?}{=} \int dz \; \ps(z|x) \nu(z) \notag \\
    &= \int dz \; \ps(z|x) \int dx' \; \ps(x'|z) \; \omega(x') \notag \\
    &= \int dx' dz \; \ps(x'|z) \; \ps(z|x) \; \omega(x') \notag \\
    \Leftrightarrow \qquad
    \int dz p(x'|z)\,p(z|x) &= \delta(x'-x) \; .
\end{align}
The unfolding only closes for a completely invertible detector.
For stochastic detectors, the OmniFold weight from
Eq.\,\eqref{eq:app-omnifold-optimum} is in a sense double-smeared, and
cannot completely map $\ps(z)$ onto $\pu(z)$. To account for this,
OmniFold iterates the reweighting and regression to converge on a
solution. By viewing these iterations as expectation-maximization
steps, the authors of OmniFold showed that it is guaranteed to
converge to a solution of Eq.\,\eqref{eq:weight-map}.

%%%%%%%%%%%%%%%%%%%%%%%%%%%%%%%%%%%%%%%%%%%%%%%%%%%%%%%%%
\section{Results}
\label{sec:res}

We benchmark the performance of AUSSIE on a set of increasingly
complex unfolding problems. We start with a Gaussian toy example in
Section~\ref{sec:res_toy} before unfolding detector effects on jets
via substructure observables in  Section~\ref{sec:res_sub} and via
the full constituent-level phase space in Section~\ref{sec:res_full}.
Finally, we show parton-level unfolding for a $tHj$ process in
Section~\ref{sec:res_mem}.

%%%%%%%%%%%%%%%%%%%%%%%%%%%%%%%%%%%%%%%%%%%%%%%%%%%%%%%%%
\subsection{Toy example}
\label{sec:res_toy}

We begin with a simple Gaussian example where the forward map imparts
a large smearing,
\begin{align}
    \begin{array}{r}
        \ps(z) = \mathcal{N}(z\,;\phantom{0.}\,0, \phantom{0.}1)\\
        \pd(z) = \mathcal{N}(z\,;\,0.2, 0.9)\\
    \end{array}
    \qqquad\text{and}\qqquad
    p(x|z) = \mathcal{N}(x\,;\,z, 2)\;.
    \label{eq:toy_data}
\end{align}
For this one-dimensional problem, we train AUSSIE using small MLP
networks for $R_\theta$ and $\Rbar_\varphi$ and the Gaussian kernel
loss from Eq.\,\eqref{eq:gauss-loss}. We make no attempt to optimize the
hyperparameter $\Lambda$ and simply set it to one. For comparison, we
also train OmniFold using the same network configurations for up to
20 iterations.

%-------------------------------------------------------
\begin{figure}[b!]
    \includegraphics[width=0.475\textwidth,
    page=1]{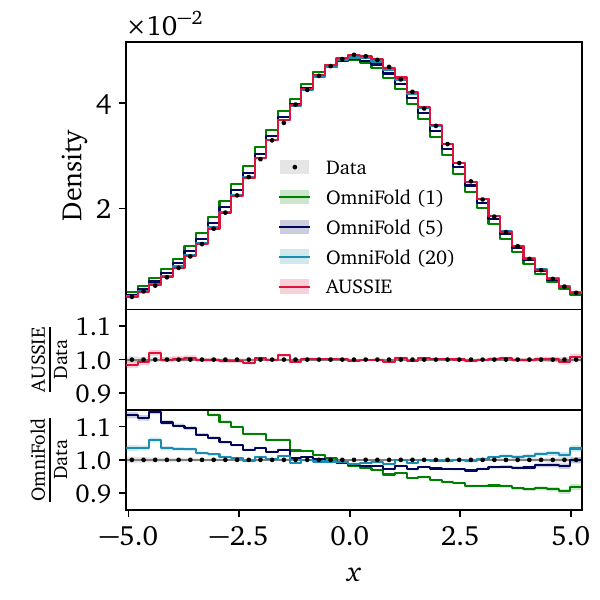}
    \includegraphics[width=0.475\textwidth]{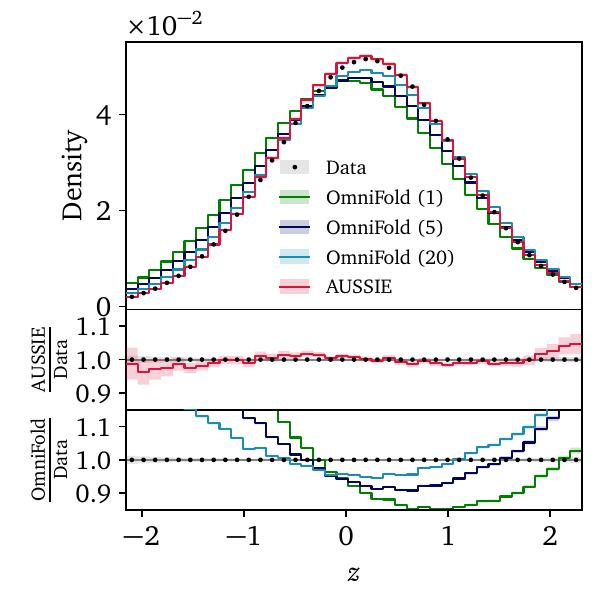}
    \caption{\textbf{Gaussian toy example:} Observable feature $x$
        (left) and unfolded feature $z$ (right), comparing AUSSIE and
    various iterations of OmniFold.}
    \label{fig:gaussian_toy}
\end{figure}
%-------------------------------------------------------

In Figure~\ref{fig:gaussian_toy} we show the results of the unfolding
for reco level (left) and part level (right). AUSSIE achieves perfect
closure on reco level and near perfect
closure on part level. The solutions on part level are somewhat
degenerate due to the finite statistics of the training sample.
Comparing this result to a single iteration of OmniFold, the
misspecified objective in Eq.\,\eqref{eq:app-omnifold-optimum}
manifests as a poor reweighting on the part and reco levels. OmniFold
improves with iterations, but the large smearing means that 20
iterations are not quite sufficient for convergence.

%-------------------------------------------------------
\begin{figure}[t]
    \includegraphics[width=\textwidth]{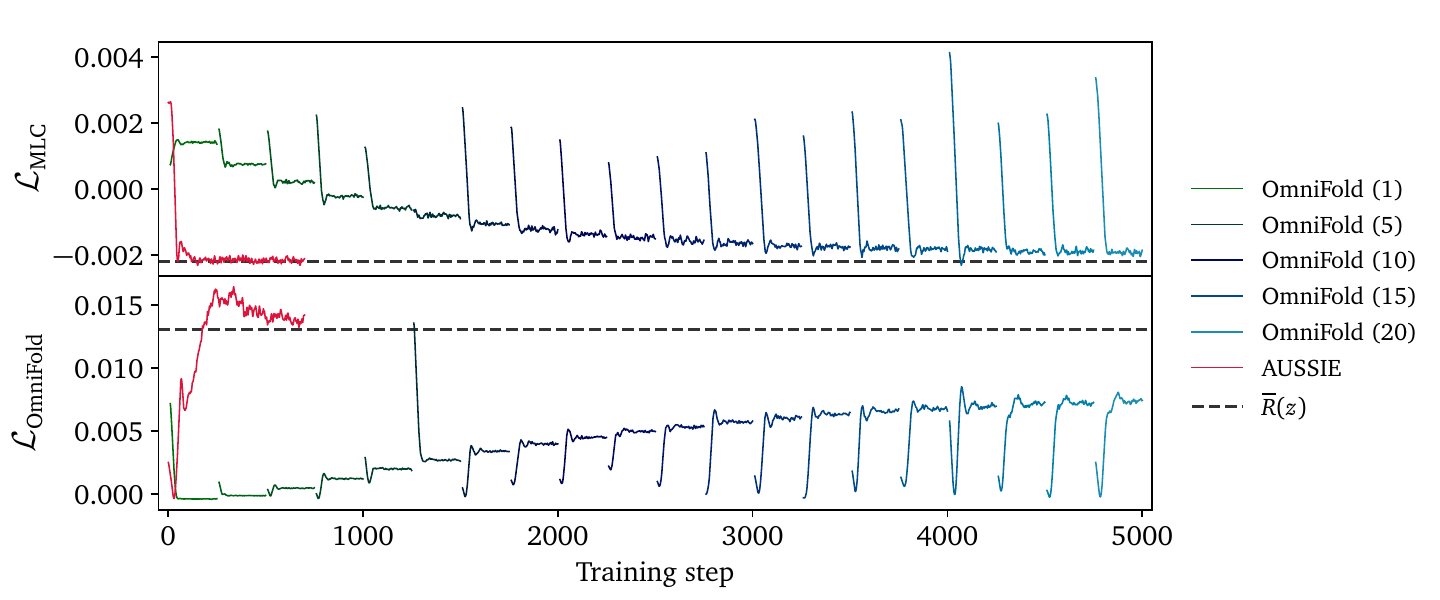}%
    \caption{\textbf{Regression losses tracked over training:} (Top)
        The $\loss_\text{MLC}$ loss, whose optimum enforces the unfolding
        condition. (Bottom) The $\loss_\text{OmniFold}$ loss, which is an
        inversion of $\loss_\text{MLC}$. The black dashed line shows the
        value of each loss for the optimal solution $\Rbar(z)$. Note that
        $\loss_\text{OmniFold}$ is only the direct objective for the
    first OmniFold iteration.}
    \label{fig:losses_chained}
\end{figure}
%-------------------------------------------------------

To highlight the different training dynamics of AUSSIE and OmniFold,
we trace their convergence with respect to the analytic solution of
our toy problem. Using Eq.\,\eqref{eq:toy_data}, we determine the true
functions $R(x)$ and $\Rbar(z)$ and calculate the corresponding
values of $\loss_\text{MLC}$ and $\loss_\text{OmniFold}$ over the
test dataset. Figure~\ref{fig:losses_chained} tracks
the loss values for the two second-step networks of AUSSIE and OmniFold. The
single AUSSIE classifier and 20 OmniFold classifiers are omitted. In
the top panel we see  that both methods approach the solution
$\Rbar(z)$, although the convergence is significantly faster for
AUSSIE. The bottom panel reveals the cause; $\Rbar(z)$ is not a
minimizer of $\loss_\text{OmniFold}$ and thus each regression step of
OmniFold spends unnecessary computation. We refrain from explicit
timing measurements in this study because the cost of OmniFold varies from task to task according to the required number of iterations, which depends on the level of smearing. A fair assessment would also require extensive optimization of the hyperparameters related to iteration.

%%%%%%%%%%%%%%%%%%%%%%%%%%%%%%%%%%%%%%%%%%%%%%%%%%%%%%%%%
\subsection{Jet substructure observables}
\label{sec:res_sub}

%-------------------------------------------------------
\begin{figure}[t]
    \includegraphics[width=0.475\textwidth, page=4]{figures/zjets_reco.pdf}
    \includegraphics[width=0.475\textwidth, page=6]{figures/zjets_reco.pdf} \\
    \includegraphics[width=0.475\textwidth, page=1]{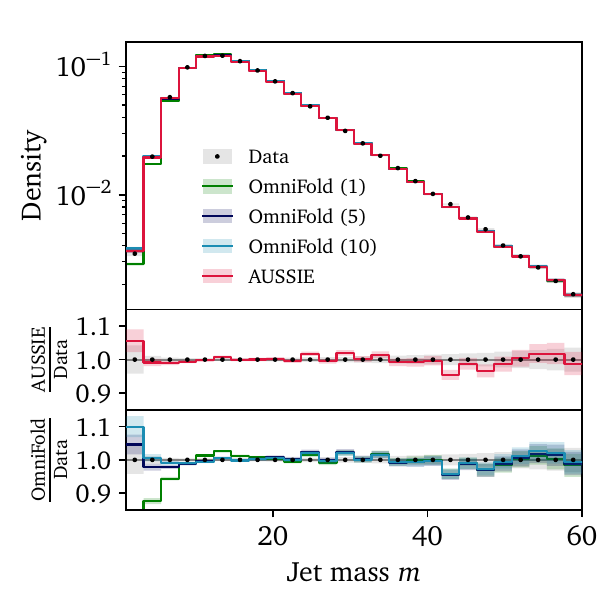}
    \includegraphics[width=0.475\textwidth, page=5]{figures/zjets_reco.pdf}
    \caption{\textbf{Reco-level jet substructure}, comparing AUSSIE and various
    iterations of OmniFold.}
    \label{fig:zjets_reco}
\end{figure}
%-------------------------------------------------------

As our first physics example, we unfold detector effects on jet
substructure observables in $Zj$ events. The standard
dataset~\cite{Andreassen:2019cjw} consists of R=0.4
${\text{anti-}k_T}$ jets with $p_T>150$~GeV that are generated with
different parton showers but passed through the same
\textsc{Delphes}~v3.4.2~\cite{deFavereau:2013fsa} simulation of the
CMS detector. The reco-level jets are defined with the same
clustering algorithm. For the reference simulation $\ps$, we take the ATLAS
A14 \textsc{Pythia}~v8.243 central tune sample, which contains 1.6M
jets. An equal-sized sample showered with  \textsc{Herwig}~v7.1.5 is
also provided. The aim is to unfold the following six jet
substructure observables
\begin{align}
    x, z =
    \begin{Bmatrix*}[r]
        {\text{Jet mass}~m}& {\text{N-subjettiness ratio}~\tau_{21}}&
        {\text{Groomed momentum fraction}~z_g} \\
        {\text{Jet width}~w}& {\text{Jet multiplicity}~N}&
        {\text{Groomed mass}~\log \rho}
    \end{Bmatrix*}\;.
\end{align}
Unfolding high-level observables induces a dependence on the
reference distribution in the results~\cite{ATLAS:2024xxl,
Butter:2025mek}. This is because the assumption that the forward
mapping in simulations matches the one in data is violated ---
transfer functions between high-level observables involve an integral
over the part-level phase space of the given distribution. In a real
measurement, this `hidden-variables' effect should be treated as a systematic
uncertainty~\cite{ATLAS:2024xxl}, or avoided by unfolding
the full particle-level phase space~\cite{Petitjean:2025tgk}. We
sidestep this issue for now and define the pseudodata as a reweighting of the
\textsc{Pythia} reference, where the weights are obtained from a
classifier trained between \textsc{Pythia} and \textsc{Herwig} on the
part-level substructure observables,
\begin{align}
    \pd(x, z) \equiv D(z)\,p_\textsc{Pythia}(x, z) \qquad \text{with}
    \qquad D(z) \approx \frac{p_\textsc{Herwig}(z)}{p_\textsc{Pythia}(z)}\;.
    \label{eq:pythia-pseudodata}
\end{align}
We train AUSSIE in this setting, again using simple MLPs and the
Gaussian Kernel loss in Eq.\,\eqref{eq:gauss-loss}. We take $\Lambda=1$ to match the variance of the input features after preprocessing, which is described in Appendix~\ref{app:hyperparams}. The effect of other kernel choices is studied in Appendix~\ref{app:metrics}. To serve as a reference,  we
also perform ten iterations of OmniFold.

We show the reco-level results for the substructure observables in
Figure~\ref{fig:zjets_reco}. AUSSIE matches the data at the order
of one percent in each observable, whereas OmniFold exhibits
a non-closure in its first iteration. After ten iterations, OmniFold
converges to closure but does not match the precision of AUSSIE for $\tau_{21}$
or $z_g$.

\begin{figure}%[t]
    \centering
    \includegraphics[width=0.475\textwidth, page=7]{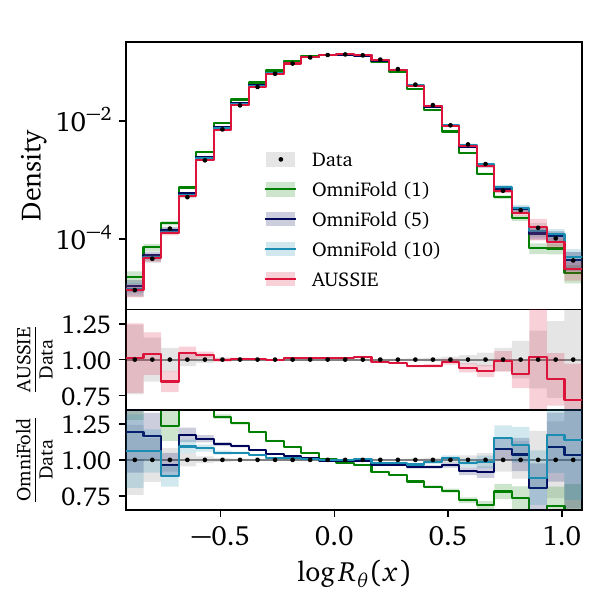}
    \vspace*{-3mm}
    \caption{\textbf{Reco-level classifier score over jet
        substructure observables},
    comparing AUSSIE and various iterations of OmniFold.}
    \label{fig:zjets_reco_class}
% \end{figure}
% %-------------------------------------------------------

% %-------------------------------------------------------
% \begin{figure}[b!]
    \includegraphics[width=0.475\textwidth, page=4]{figures/zjets_part.pdf}
    \includegraphics[width=0.475\textwidth, page=6]{figures/zjets_part.pdf} \\
    \includegraphics[width=0.475\textwidth, page=1]{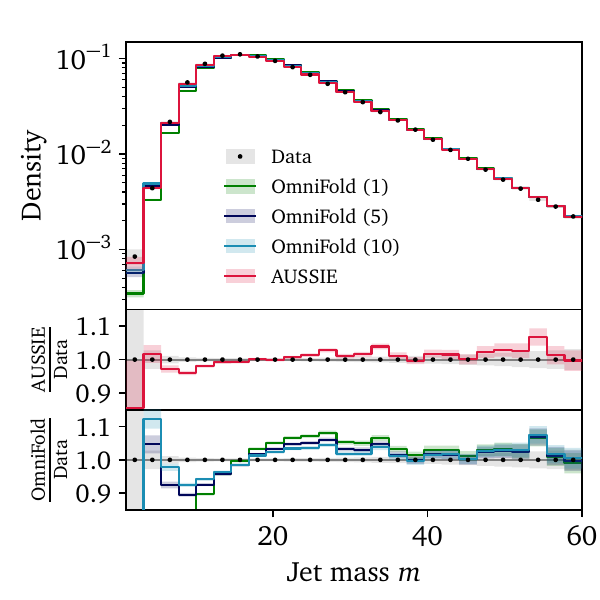}
    \includegraphics[width=0.475\textwidth, page=5]{figures/zjets_part.pdf}
    \vspace*{-3mm}
    \caption{\textbf{Unfolded part-level jet substructure}, comparing
    AUSSIE and various iterations of OmniFold.}
    \label{fig:zjets_part}
\end{figure}

In Figure~\ref{fig:zjets_reco_class} we summarize the reco-level
performance using the score of the first-step classifier
$R_\theta(x)$, which approximates the optimal observable. AUSSIE maps
the lower tail precisely while the upper tail shows a small
deviation. OmniFold also reaches relatively high precision after ten
iterations, but remains slightly biased toward the simulation as
evidenced by the excess in the lower tail.

%-------------------------------------------------------
%-------------------------------------------------------

Finally, in Figure~\ref{fig:zjets_part} we see that the
non-convergence of OmniFold is most evident on part level. It now
also shows significant deviations from truth in the jet mass, in
addition to $\tau_{21}$
and~$z_g$. AUSSIE is more consistent with truth, with deviations
being limited to the tails. Since we already observed a tight closure
for AUSSIE on reco-level, these deviations mostly reflect the
degenerate solution space. For quantitative performance metrics, we refer the reader to Appendix~\ref{app:metrics}.

% \clearpage
%%%%%%%%%%%%%%%%%%%%%%%%%%%%%%%%%%%%%%%%%%%%%%%%%%%%%%%%%
\subsection{Jet constituents}
\label{sec:res_full}

A straightforward solution to the hidden-variables problem, and the
proper unfolding task for public data, is to unfold all jet
constituents. Since the forward detector simulation naturally acts on
this phase space, the unfolding assumption of a universal transfer
function is valid, at least up to nuisance parameters. We study
AUSSIE in this setting and, accordingly,
take the $\textsc{Herwig}$ sample from the jet dataset as the
pseudodata, in contrast to Eq.\,\eqref{eq:pythia-pseudodata}. The vastly
increased dimensionality of this unfolding problem poses a
significant challenge.

We represent the part-level and reco-level jets as sets of constituent
four momenta and now use L-GATr
networks~\cite{Brehmer:2024yqw, Petitjean:2025zjf}
for $R_\theta(x)$ and $\Rbar_\varphi(z)$ to exploit the Lorentz and
permutation symmetries. For the same reason, and to handle events
with varying length, we also switch from the Gaussian kernel loss to
the AutoDiff loss from Eq.\,\eqref{eq:autodiff-loss-l1}.

%-------------------------------------------------------
\begin{figure}[b!]
    \includegraphics[width=0.475\textwidth,
    page=2]{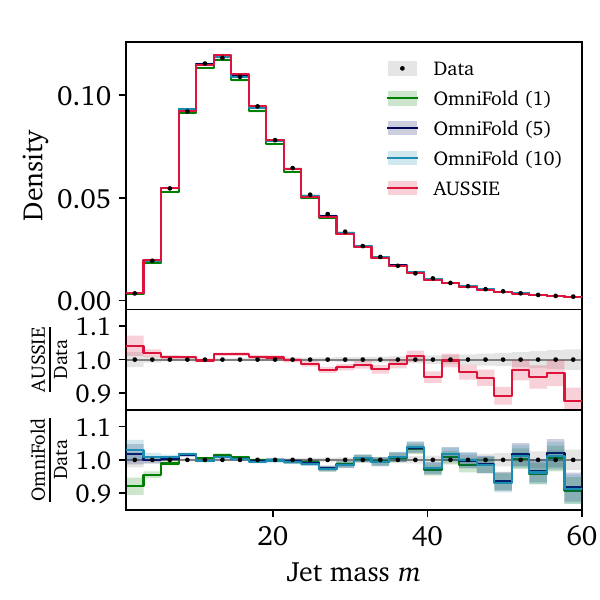}
    \includegraphics[width=0.475\textwidth,
    page=3]{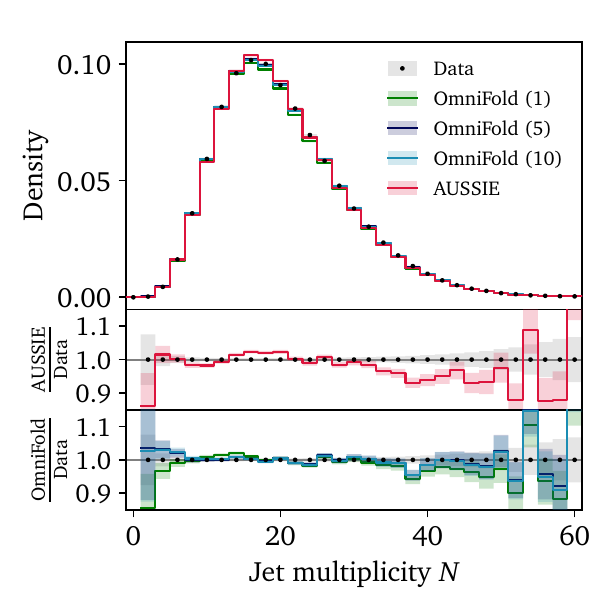} \\
    \includegraphics[width=0.475\textwidth,
    page=5]{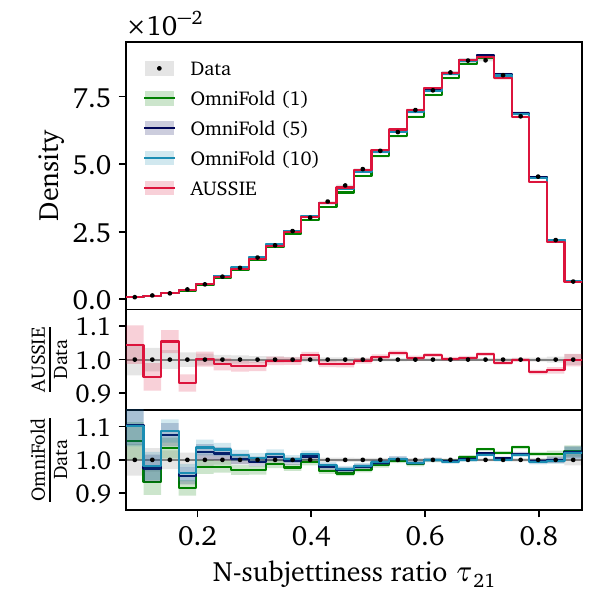}
    \includegraphics[width=0.475\textwidth,
    page=6]{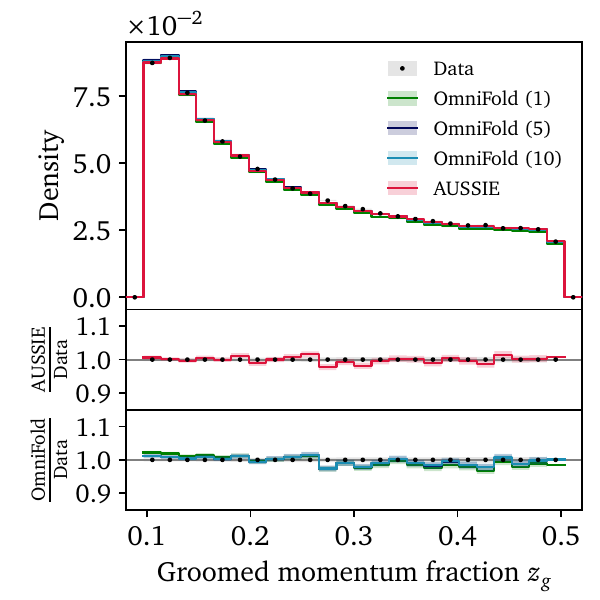}
    \caption{\textbf{Reco-level jet substructure (using
    constituents)}, comparing AUSSIE and various iterations of OmniFold.}
    \label{fig:zjets_particle_reco}
\end{figure}
%-------------------------------------------------------

\begin{figure}
    \centering
    \includegraphics[width=0.475\textwidth,
    page=8]{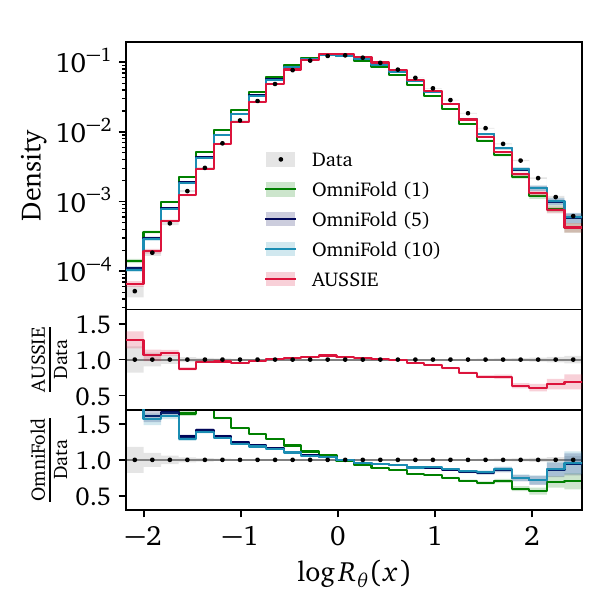}
    \vspace*{-3mm}
    \caption{\textbf{Reco-level classifier score over
        constituent-level jets}, comparing AUSSIE and various iterations
    of OmniFold.}
    \label{fig:zjets_particle_reco_class}
\end{figure}

Starting with reco-level, we show the jet substructure observables in
Figure~\ref{fig:zjets_particle_reco}, this time calculated using the
recorded constituents. The difference between AUSSIE and OmniFold at
reco-level is smaller than in the high-level case from
Figure~\ref{fig:zjets_reco}. This is because the forward map now
corresponds to the true action of the detector simulation, which is
only a modest smearing. As a result OmniFold now shows better closure
in its first iteration. AUSSIE still improves on iteration ten of
OmniFold in $\tau_{21}$ and $z_g$, but varies slightly more in the
jet mass and multiplicity. However, the classifier score in
Figure~\ref{fig:zjets_particle_reco_class} shows that these marginals
obscure information in other constituent-level correlations. While
neither AUSSIE nor OmniFold completely reweights the classifier score
to match data, AUSSIE does correct the lower tail indicating a better
closure in the full phase space. Unlike the previous result using the
high-level jet substructure observables, there is no significant
improvement for OmniFold
between iterations 5 and 10.

\begin{figure}
    \centering
    %-------------------------------------------------------
    \includegraphics[width=0.475\textwidth,
    page=2]{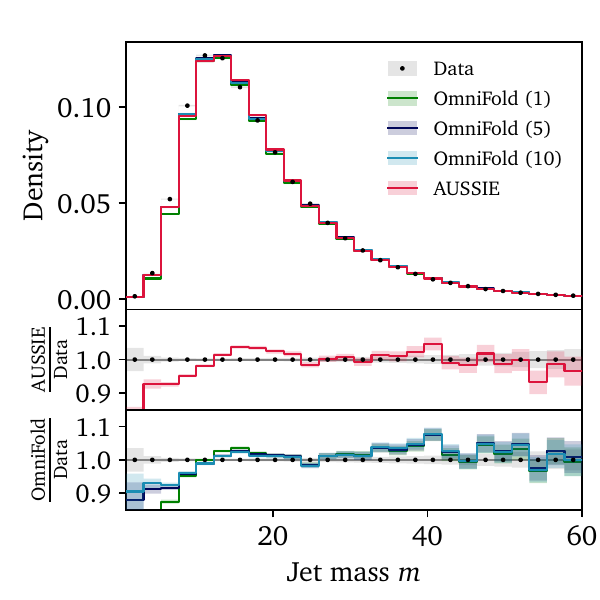}
    \includegraphics[width=0.475\textwidth,
    page=3]{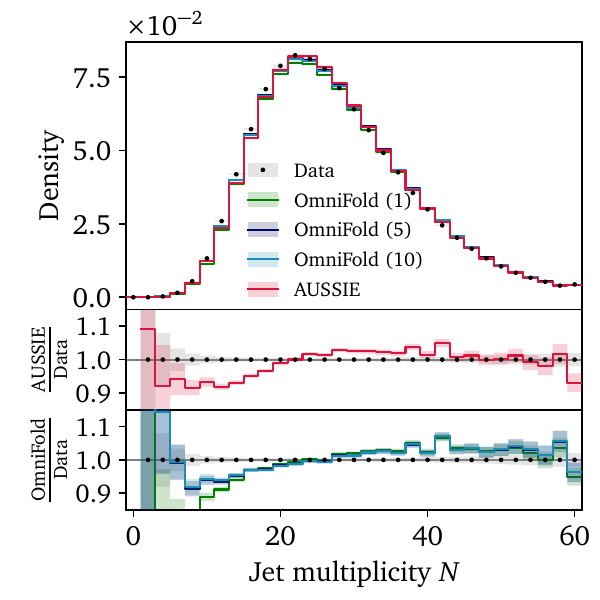} \\
    \includegraphics[width=0.475\textwidth,
    page=5]{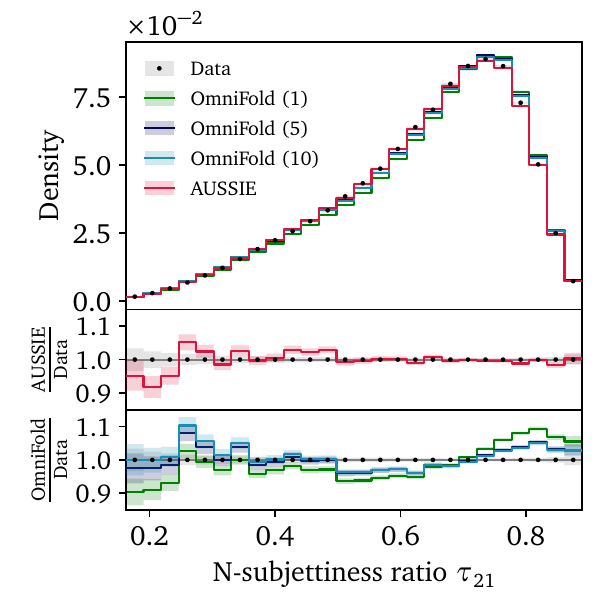}
    \includegraphics[width=0.475\textwidth,
    page=6]{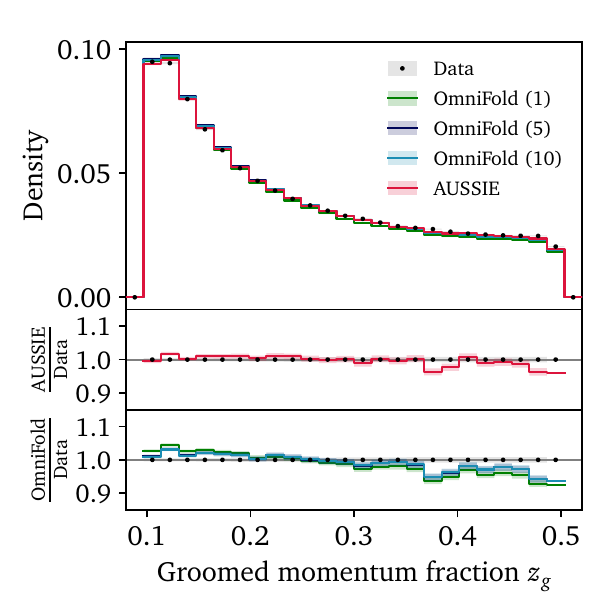}
    \vspace*{-3mm}
    \caption{\textbf{Unfolded part-level jet substructure (using
    constituents)}, comparing AUSSIE and various iterations of OmniFold.}
    \label{fig:zjets_particle_part}
\end{figure}

On part-level, in Figure~\ref{fig:zjets_particle_part}, we find that
neither approach closes exactly onto the \textsc{Herwig} truth. The
jet mass and multiplicity show deviations on the order of ten percent
for both methods. Still, $\tau_{21}$ and $z_g$ are recovered well by
AUSSIE while the OmniFold iterations stagnate with larger bias. As
discussed earlier, one cannot expect the part-level distributions to
match while observing non-closure at reco level. In this case, we
find that the high dimensionality is even a challenge for the
step-one classifier. Both
AUSSIE and OmniFold are likely to benefit from a larger dataset or
transfer learning from
pretrained networks~\cite{Bhimji:2025isp, Mikuni:2025tar, Mikuni:2024qsr}.

%-------------------------------------------------------
%-------------------------------------------------------
%-------------------------------------------------------
%-------------------------------------------------------

\clearpage
%%%%%%%%%%%%%%%%%%%%%%%%%%%%%%%%%%%%%%%%%%%%%%%%%%%%%%%%%
\subsection{Parton-level events}
\label{sec:res_mem}

As a final example, we showcase AUSSIE for unfolding to parton level,
which is relevant for parameter inference with the ML-Matrix Element
Method~\cite{Butter:2022vkj, Heimel:2023mvw}. Reweighting-based unfolding
is well suited to this scenario since it automatically preserves
strict correlations such as momentum conservation and on-shell
conditions. However, as we will shortly demonstrate, the high degree
of information loss in this setting poses a challenge to iterative
methods such as OmniFold.

We consider associated single-top and Higgs
production~\cite{Butter:2022vkj, Heimel:2023mvw}
\begin{align}
    pp \to tHj \to (bjj)\, (\gamma \gamma)\;j + \text{ISR}\;,
\end{align}
where the top quark decays hadronically and the Higgs decays to
photons. ISR introduces additional jets at reco level, effectively
causing a combinatorial smearing that prevents clean reconstruction
of the top. This process provides sensitivity to the CP phase in the
top-Yukawa coupling and is a prime use case for machine learning due
to its small cross section in the Standard Model.

We use the dataset from Ref.~\cite{Butter:2022vkj}, which considers a
varying CP angle $\alpha\in(-\pi,\pi]$, defined through the Yukawa
\begin{align}
    \mathscr{L}_{t\bar tH} = -\frac{y_t}{\sqrt{2}} \left[\cos \alpha
    \,t\bar t + \frac{2i}{3}\sin\alpha\,t\gamma_5\bar t\right] H\;.
\end{align}
The reference simulation samples $\alpha$ uniformly while the
pseudodata fixes the angle,
\begin{align}
    \ps(z) =
    \XXLangle\frac{1}{\sigma(\alpha)}\frac{d\sigma(\alpha)}{dz}\XXRangle_{\alpha\sim\mathcal{U}[-\pi,\pi]}\;,
    \qquad     \pd(z) =
    \left.\frac{1}{\sigma(\alpha)}\frac{d\sigma(\alpha)}{dz}\right|_{\alpha=\pi/4}\;.
\end{align}
The parton-level events are generated by \textsc{Madgraph5}~v3.1.0
with LO-NNPDF, showered with \textsc{Pythia}, and passed to
\textsc{Delphes} for detector simulation and reconstruction of
$R=0.4$ anti-$k_T$ jets. We keep at most four ISR jets in the
reco events, for a maximum of ten total objects. The
dataset contains 6.8M reference simulation events and 440k pseudodata
events. For more specifics on the process we refer the reader to
Ref.~\cite{Butter:2022vkj}.

\begin{figure}[t]
    \includegraphics[width=0.475\textwidth, page=2]{figures/yukawa_reco.pdf}
    \includegraphics[width=0.475\textwidth, page=1]{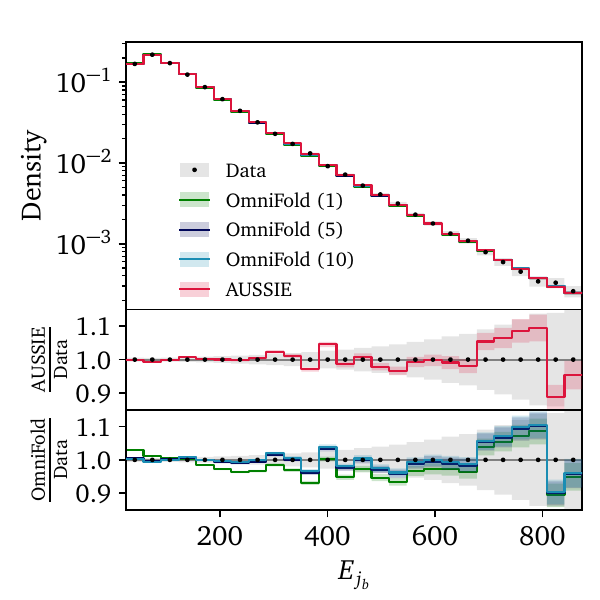}  \\
    \includegraphics[width=0.475\textwidth, page=3]{figures/yukawa_reco.pdf}
    \includegraphics[width=0.475\textwidth, page=4]{figures/yukawa_reco.pdf}
    \caption{\textbf{Reco-level observables in $\bm{tHj}$ events},
        comparing AUSSIE and various iterations of
    OmniFold. The lower-right panel is the first-step classifier score.}
    \label{fig:yukawa_reco}
\end{figure}

Just as in the constituent unfolding above, here we apply L-GATr
networks and the AutoDiff loss to the full part-level and reco-level
phase spaces.
The reco-level results with AUSSIE and OmniFold are shown in
Figure~\ref{fig:yukawa_reco}. In addition to three kinematic
variables that are sensitive to the CP angle, we also include the
first-step classifier approximation of the optimal
reco-level observable. AUSSIE
produces perfect closure in all of these variables, including the
classifier score. Even after iterating, OmniFold retains deviations
in the tails of the kinematic variables that become apparent in the
classifier score. As in the previous section, OmniFold does not
noticeably improve from iteration five and ten. We have also checked that no
intermediate iteration matches AUSSIE.

\begin{figure}[t]
    \includegraphics[width=0.475\textwidth, page=4]{figures/yukawa_part.pdf}
    \includegraphics[width=0.475\textwidth, page=1]{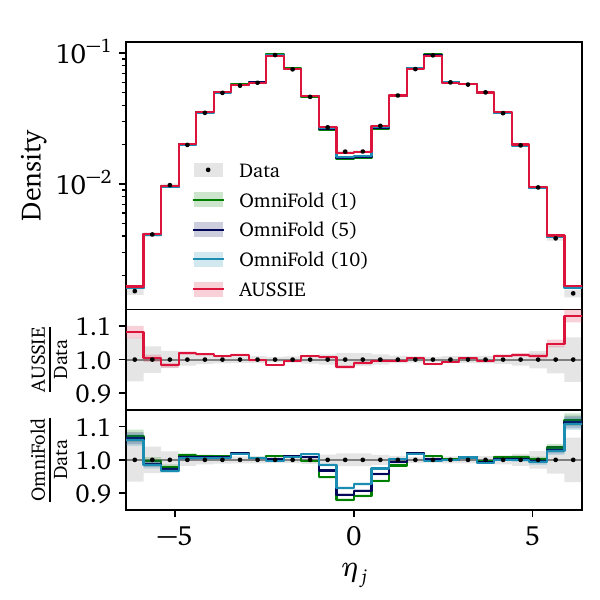}  \\
    \includegraphics[width=0.475\textwidth, page=3]{figures/yukawa_part.pdf}
    \includegraphics[width=0.475\textwidth, page=2]{figures/yukawa_part.pdf}
    \vspace*{-3mm}
    \caption{\textbf{Unfolded parton-level features in $\bm{tHj}$
    events}, comparing AUSSIE and various iterations of OmniFold.}
    \label{fig:yukawa_part}
\end{figure}

We show a selection of part-level distributions in
Figure~\ref{fig:yukawa_part}. As we have seen before, the advantage
of AUSSIE is magnified for part level, compared to reco level.
Specifically, angular correlations are
reproduced to within statistical fluctuations by AUSSIE, even in
regions where OmniFold fails to converge. While OmniFold and AUSSIE
both perform well for this benchmark process, we find no observables in
which AUSSIE does not improve at least slightly.

%-------------------------------------------------------
%-------------------------------------------------------

%-------------------------------------------------------
%-------------------------------------------------------

\clearpage
%%%%%%%%%%%%%%%%%%%%%%%%%%%%%%%%%%%%%%%%%%%%%%%%%%%%%%%%%
\section{Outlook}
\label{sec:outlook}
We introduced Adversary-free Unfolding SanS Iteration or
Emulation (AUSSIE), a new method for unbinned unfolding of
multidimensional collider observables. AUSSIE is a two-step algorithm
based on reweighting a simulated dataset that numerically inverts the
integral equation relating latent and observable phase spaces.
Importantly, it is non-iterative and does not suffer from
explicit dependence on the simulated dataset. To our knowledge,
AUSSIE is the only unfolding method that achieves this without
relying on surrogates or adversarial training.

We demonstrated the performance of AUSSIE on three physics
scenarios: unfolding detector effects on jets at the level of
substructure observables and at the level of constituents, as well as
complete event
unfolding to parton level. AUSSIE consistently yields more accurate
unfolded distributions than ten iterations of OmniFold, showing
better closure in both the reco-level and part-level phase spaces.

For experimental effects such as irreducible backgrounds and imperfect
acceptance or efficiency, AUSSIE can be readily combined with existing
techniques~\cite{Falcao:2025jom, Butter:2025mek}. The AUSSIE weights can also be used to define an informed prior for methods that learn a generative inverse map, as discussed in~\cite{Vandegar:2020yvw}.

While we have showcased AUSSIE for the task of unfolding, it is more
generally a tool for efficient simulation-based inference which we expect
to apply more broadly.

%%%%%%%%%%%%%%%%%%%%%%%%%%%%%%%%%%%%%%%%%%%%%%%%%%%%%%%%%
\subsection*{Acknowledgments}

We acknowledge support by the Deutsche Forschungsgemeinschaft (DFG,
German Research Foundation) under grant 396021762 -- TRR~257:
\textsl{Particle Physics Phenomenology after the Higgs Discovery},
and through Germany's Excellence Strategy EXC~2181/1 -- 390900948
(the \textsl{Heidelberg STRUCTURES Excellence Cluster}). We
acknowledge support by the state of Baden-Württemberg through bwHPC
and the German Research Foundation (DFG) through grant INST 35/1597-1 FUGG. This
project started at the Kavli Institute for Theoretical
Physics (KITP) supported by grant NSF PHY-2309135.

%%%%%%%%%%%%%%%%%%%%%%%%%%%%%%%%%%%%%%%%%%%%%%%%%%%%%%%%%
\subsection*{Code availability}
The code for AUSSIE can be found at
\url{https://github.com/heidelberg-hepml/aussie}.

\appendix
%%%%%%%%%%%%%%%%%%%%%%%%%%%%%%%%%%%%%%%%%%%%%%%%%%%%%%%%%
\section{Training details and hyperparameters}
\label{app:hyperparams}

\subsection*{Toy example}
For the Gaussian toy example from Section~\ref{sec:res_toy}, we do
not use any preprocessing. The full set of network and optimization
hyperparameters are listed in Table~\ref{tab:hyperparams_toy}.

\subsection*{Jet substructure}
As preprocessing for the substructure observables in the jet dataset
from Section~\ref{sec:res_sub}, we first apply a logarithmic scaling
at both part and reco level:
\begin{align}
    m \to \log m, \quad \tau_{21} \to \log \tau_{21}, \quad
    N\to\log{N}, \quad w \to \log w\;.
\end{align}
All features are then standardized to zero mean and unit variance,
using separate transformations for $x$ and $z$ calculated with the
training split of the reference simulation. The full set of network
and optimization hyperparameters are listed in
Table~\ref{tab:hyperparams_zjet}. The network used to train $D(z)$ as
defined in Eq.\,\eqref{eq:pythia-pseudodata} shares the same MLP
architecture as the classifier and unfolder.

\subsection*{Constituent-level jets}
For the jet constituent example in Section~\ref{sec:res_full}, we
apply preprocessing compatible with the equivariance constraints of
the L-GATr architecture. All four momenta are represented in cartesian
coordinates $(E, p_x, p_y, p_z)$ and preprocessed by boosting into
the center-of-mass frame of the jet. Particle identities are one-hot
encoded and enter the network as scalar channels. To allow for
symmetry breaking, reference vectors indicating the time and
beam-transverse directions,
\begin{align}
    \big\{(1, 0, 0, 0),\, (1, 0, 0, 1),\, (1, 0, 0, -1)\big\},
    \label{eq:ref-vectors}
\end{align}
are appended to the collection of particles before preprocessing with
a unique scalar identifier.
The full set of network and optimization hyperparameters are listed
in Table~\ref{tab:hyperparams_zjet_particle}.

\subsection*{Parton-level events}

For the $tHj$ example in Section~\ref{sec:res_mem}, we also use
Lorentz-equivariant preprocessing. However, instead of boosting
events we globally scale the four-momenta by 1/450~GeV and 1/140~GeV
at the part and reco levels respectively. All reco light jets share
the same scalar particle identifier, and we add the same
symmetry-breaking reference vectors from Eq.\,\eqref{eq:ref-vectors}.
The full set of network and optimization hyperparameters are listed
in Table~\ref{tab:hyperparams_yukawa}.

\begin{table}
    \centering
    \begin{small}
        \begin{tabular}{lcc}
            \toprule
            & Classifier $R_\theta$   & Unfolder $\Rbar_\varphi$  \\
            \midrule
            MLP hidden channels         & 2            & 2                \\
            MLP hidden layers           & 32           & 32               \\
            MLP activation              & SiLU         & SiLU             \\
            \midrule
            $N_\text{data}$             & 10M          & -                \\
            $N_\text{sim}$              & 10M          & 10M              \\
            Train/Val/Test              & 60/5/35      & 60/5/35          \\
            Optimizer                        & ScheduleFree AdamW &
            ScheduleFree AdamW\\
            Learning rate               & $5\times10^{-3}$    &
            $5\times10^{-3}$ \\
            Batch size                  & 8192         & 8192             \\
            Epochs                      & 1            & 1                \\
            Weight decay                & 0            & 0                \\
            Warmup steps                & 200          & 200              \\
            \bottomrule
        \end{tabular}
    \end{small}
    \caption{Network and optimization hyperparameters for AUSSIE and
    OmniFold on the Gaussian toy example from Section~\ref{sec:res_toy}.}
    \label{tab:hyperparams_toy}
\end{table}
\begin{table}
    \centering
    \begin{small}
        \begin{tabular}{lcc}
            \toprule
            & Classifier $R_\theta$   & Unfolder $\Rbar_\varphi$  \\
            \midrule
            Architecture            & MLP          & MLP               \\
            Hidden channels         & 128           & 128               \\
            Hidden layers           & 4            & 4                \\
            Activation              & SiLU         & SiLU             \\
            \midrule
            $N_\text{data}$             & 1.6M          & -                \\
            $N_\text{sim}$              & 1.6M          & 1.6M              \\
            Train/Val/Test              & 60/5/35      & 60/5/35          \\
            Optimizer                        & ScheduleFree AdamW &
            ScheduleFree AdamW\\
            Learning rate               & $5\times10^{-4}$    & $10^{-3}$ \\
            Batch size                  & 1024         & 8192             \\
            Epochs                      & 200          & 500              \\
            Patience                    & 50           & 50               \\
            Weight decay                & 0            & 0                \\
            Warmup steps                & 5000         & 5000             \\
            \bottomrule
        \end{tabular}
    \end{small}
    \caption{Network and optimization hyperparameters for AUSSIE and
        OmniFold on the substructure-level jet dataset from
    Section~\ref{sec:res_sub}.}
    \label{tab:hyperparams_zjet}
\end{table}
\begin{table}
    \centering
    \begin{small}
        \begin{tabular}{lcc}
            \toprule
            & Classifier $R_\theta$ & Unfolder $\Rbar_\varphi$  \\
            \midrule
            Architecture            & L-GATr Slim & L-GATr Slim             \\
            Blocks                  & 4          & 4                      \\
            Heads                   & 4          & 4                      \\
            Scalar channels         & 64         & 64                     \\
            Vector channels         & 32         & 16                     \\
            MLP ratio               & 4          & 4                      \\
            Activation              & GELU       & GELU                   \\
            Dropout                 & 0.01       & 0.01                   \\
            \midrule
            $N_\text{data}$             & 1.6M          & -               \\
            $N_\text{sim}$              & 1.6M          & 1.6M            \\
            Train/Val/Test              & 60/5/35      & 60/5/35          \\
            Optimizer                        & ScheduleFree AdamW &
            ScheduleFree AdamW\\
            Learning rate               & $10^{-3}$    & $5\times10^{-3}$ \\
            Batch size                  & 256          & 2048             \\
            Epochs                      & 500          & 500              \\
            Patience                    & 50           & 50               \\
            Weight decay                & 0            & 0                \\
            Warmup steps                & 5000         & 5000             \\
            \bottomrule
        \end{tabular}
    \end{small}
    \caption{Network and optimization hyperparameters for AUSSIE and
        OmniFold on the constituent-level jet dataset from
    Section~\ref{sec:res_sub}.}
    \label{tab:hyperparams_zjet_particle}
\end{table}
\begin{table}
    \centering
    \begin{small}
        \begin{tabular}{lcc}
            \toprule
            & Classifier $R_\theta$ & Unfolder $\Rbar_\varphi$  \\
            \midrule
            Architecture            & L-GATr Slim & L-GATr Slim       \\
            Blocks                  & 4          & 4                \\
            Heads                   & 4          & 4                \\
            Scalar channels         & 32         & 32               \\
            Vector channels         & 8          & 8                \\
            MLP ratio               & 2          & 2                \\
            Activation              & GELU       & GELU             \\
            \midrule
            $N_\text{data}$             & 440k         & -                \\
            $N_\text{sim}$              & 6.8M         & 6.8M             \\
            Train/Val/Test              & 60/5/35      & 60/5/35          \\
            Optimizer                        & ScheduleFree AdamW &
            ScheduleFree AdamW\\
            Learning rate               & $10^{-3}$    & $5\times10^{-3}$ \\
            Batch size                  & 512          & 8192             \\
            Epochs                      & 500          & 500              \\
            Patience                    & 50           & 50               \\
            Weight decay                & 0            & 0                \\
            Warmup steps                & 5000         & 5000             \\
            \bottomrule
        \end{tabular}
    \end{small}
    \caption{Network and optimization hyperparameters for AUSSIE and
    OmniFold on the $tHj$ dataset from Section~\ref{sec:res_mem}.}
    \label{tab:hyperparams_yukawa}
\end{table}
\clearpage
%%%%%%%%%%%%%%%%%%%%%%%%%%%%%%%%%%%%%%%%%%%%%%%%%%%%%%%%%
\section{Numerical details}
\label{app:numerics}

Here we give details on the numerical estimation of the
two losses described in the main text.

\textbf{Kernel loss:} The loss from Eq.\,\eqref{eq:gauss-loss} can be
estimated by coupling paired batch elements via a kernel matrix,
\begin{align}
    \loss_\text{Gauss}\approx\frac{1}{M(M-1)}\sum_{i,j=1}^M
    \left[\left(1-\frac{\Rbar_\varphi(z_i)}{R_\theta(x_i)}\right)
        K(x_i,x_j)(1-\delta_{ij})
    \left(1-\frac{\Rbar_\varphi(z_j)}{R_\theta(x_j)}\right)\right]\;.
\end{align}
By zeroing the diagonal of the matrix, this `U-statistic' estimator
is unbiased. The resulting squared norm can occasionally take small
negative values due to numerical precision, so we clamp it to
stabilize optimization.

\textbf{AutoDiff loss:} The squared $L_2$ norm from
Eq.\,\eqref{eq:autodiff-loss-l2} can be directly computed over a
batch by evaluating
\begin{align}
    \left\lvert \nabla_\theta \loss_\text{MLC} \right\rvert^2
    \approx\left\lvert\frac{1}{M}\nabla_\theta\sum_{i=1}^M
    \ell(x_i,z_i) \right\rvert^2 \quad \text{with}\quad
    \ell(x,z)\equiv R_\theta(x) - \Rbar_\varphi(z)\log R_\theta(x)\;.
\end{align}
While this estimator is straightforward to implement with automatic
differentiation, it is biased at finite batch size: the square
induces a strictly positive contribution from gradient noise
proportional to its variance. This additive variance term vanishes in
the limit of large batches. This also explains why we find better
performance with the $L_1$ alternative from
Eq.\,\eqref{eq:autodiff-loss-l1}, since its linear scaling suppresses
this bias compared to the quadratic of the $L_2$.

An unbiased estimator for the squared $L_2$ norm can be constructed
by partitioning the batch into two (or more) disjoint subsets and
forming the inner product of their parameter gradients.
\begin{align}
    \left\lvert \nabla_\theta \loss_\text{MLC}
    \right\rvert^2\approx\Bigg[\frac{1}{P}\nabla_\theta \sum_{i=1}^P
        \ell(x_i,z_i)
    \Bigg]^T\Bigg[\frac{1}{M-P}\nabla_\theta\hspace{-1mm}\sum_{j=P+1}^M
    \ell(x_j,z_j) \Bigg]\;.
\end{align}
This two-sample estimator is consistent and removes the additive bias
from gradient variance. However, it can fluctuate below zero, making
stable optimization difficult. As in the kernel case, larger batch
sizes significantly reduce this effect. The optimization can be
stabilized by clamping the norm at the cost of some bias. For our
studies, we found the $L_1$ norm from
Eq.\,\eqref{eq:autodiff-loss-l1} more favorable, though this may
differ depending on the application.

%%%%%%%%%%%%%%%%%%%%%%%%%%%%%%%%%%%%%%%%%%%%%%%%%%%%%%%%%
\section{Additional jet substructure plots}
\label{app:additional-plots}

\begin{figure}
    \includegraphics[width=0.475\textwidth,
    page=2]{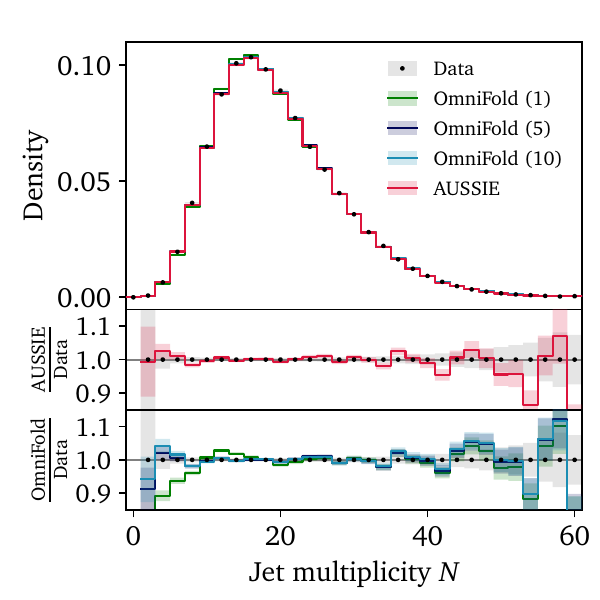}
    \includegraphics[width=0.475\textwidth,
    page=3]{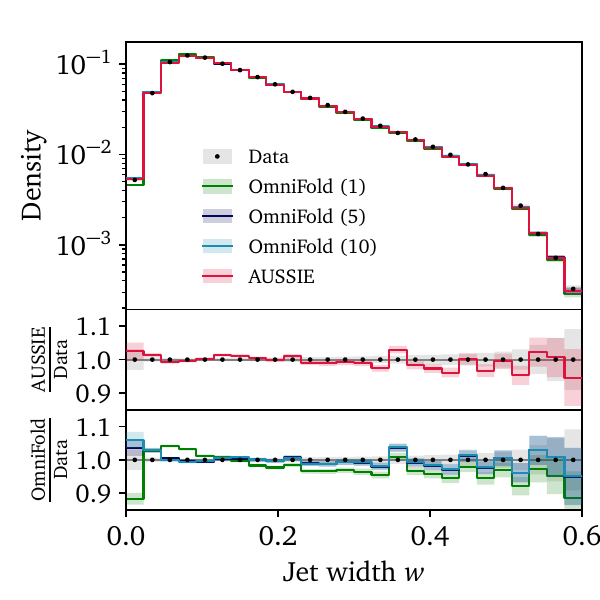}
    \vspace*{-3mm}
    \caption{Reco-level jet substructure, comparing AUSSIE and
    various iterations of OmniFold. Supplements Figure~\ref{fig:zjets_reco}.}
    \label{fig:zjets_reco_add}
\end{figure}

\begin{figure}
    \includegraphics[width=0.475\textwidth,
    page=2]{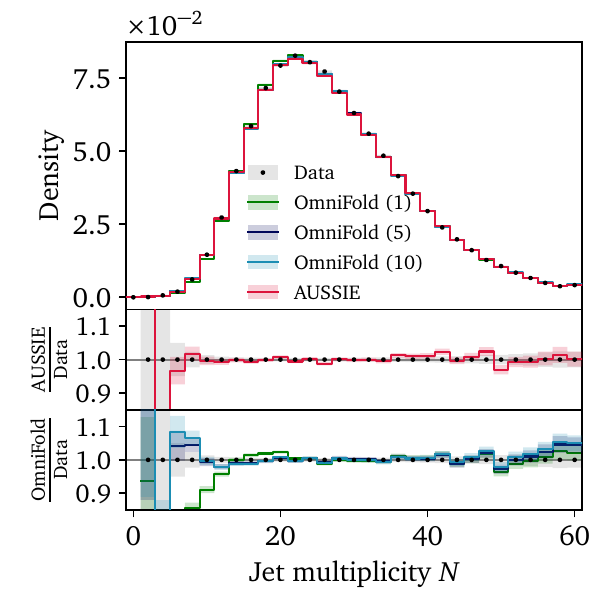}
    \includegraphics[width=0.475\textwidth,
    page=3]{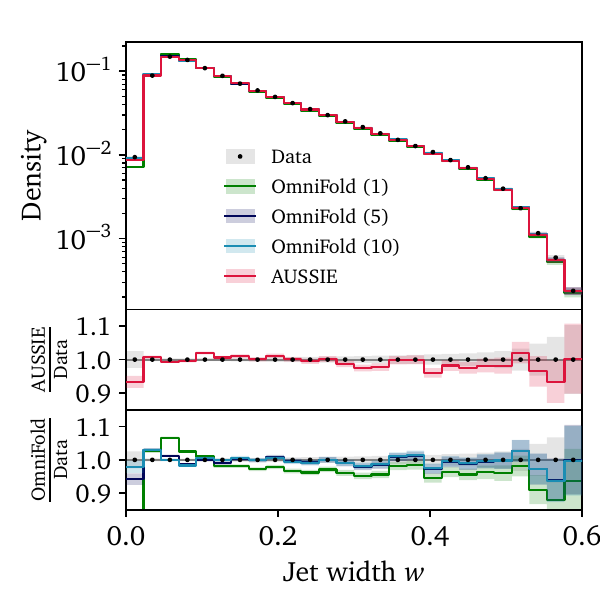}
    \vspace*{-3mm}
    \caption{Unfolded part-level jet substructure, comparing AUSSIE
        and various iterations of OmniFold. Supplements
    Figure~\ref{fig:zjets_part}.}
    \label{fig:zjets_part_add}
\end{figure}

\begin{figure}
    \includegraphics[width=0.475\textwidth,
    page=4]{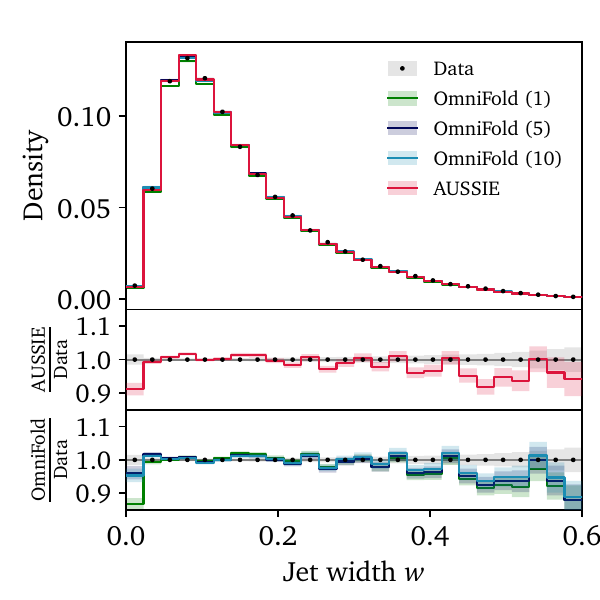}
    \includegraphics[width=0.475\textwidth,
    page=7]{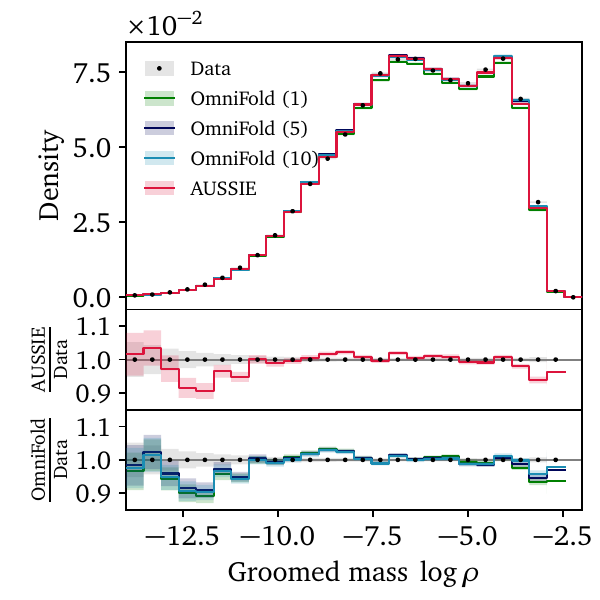}
    \vspace*{-3mm}
    \caption{Reco-level jet substructure (using constituents),
        comparing AUSSIE and various iterations of OmniFold. Supplements
    Figure~\ref{fig:zjets_particle_reco}.}
    \label{fig:zjets_particle_reco_add}
\end{figure}

\begin{figure}
    \includegraphics[width=0.475\textwidth,
    page=4]{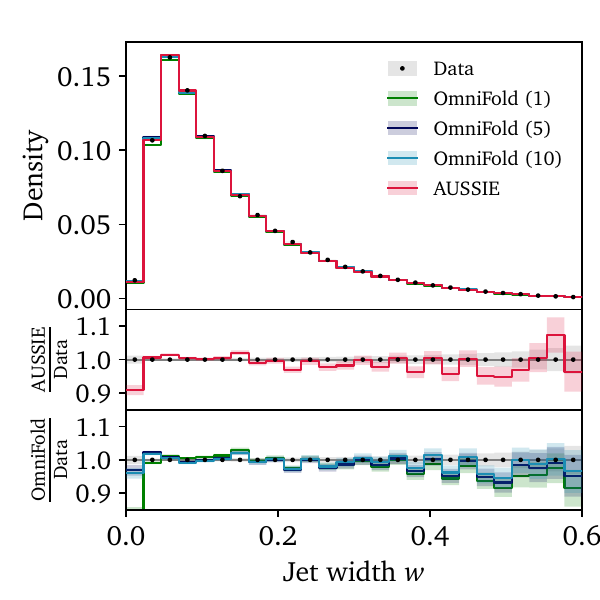}
    \includegraphics[width=0.475\textwidth,
    page=7]{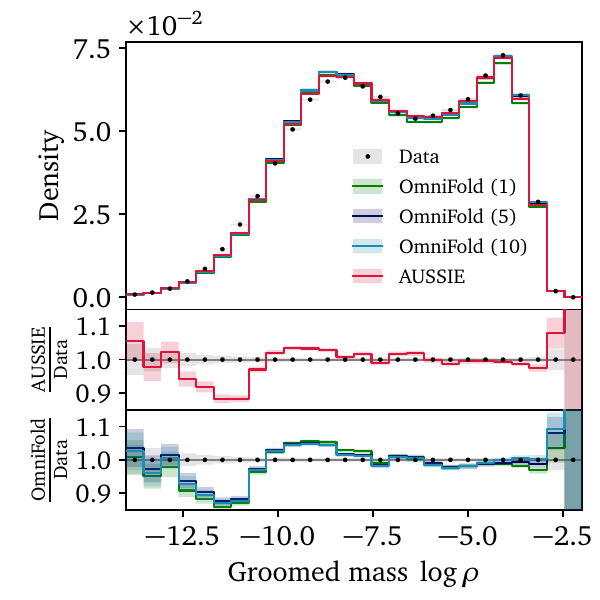}
    \vspace*{-3mm}
    \caption{Unfolded part-level jet substructure (using
        constituents), comparing AUSSIE and various iterations of
    OmniFold. Supplements Figure~\ref{fig:zjets_particle_part}.}
    \label{fig:zjets_particle_part_add}
\end{figure}

For completeness, we provide the reco-level and part-level
distributions for the substructure observables that were omitted from
the main text. Figures~\ref{fig:zjets_reco_add} and
\ref{fig:zjets_part_add} respectively show the reco-level and
part-level results for the jet multiplicity and width using the
high-level dataset. Figures~\ref{fig:zjets_particle_reco_add} and
\ref{fig:zjets_particle_part_add} the results for the jet
multiplicity and groomed mass.


%%%%%%%%%%%%%%%%%%%%%%%%%%%%%%%%%%%%%%%%%%%%%%%%%%%%%%%%%
\clearpage
\section{Kernel dependence and \texorpdfstring{$\chi^2$}{Chi2} metrics}
\label{app:metrics}
Here we supplement the results plots from the main text with quantitative performance metrics and study the dependence on the choice of kernel. In addition to the Gaussian kernel from Eq.\,\ref{eq:kernel-gauss}, we consider the exponential kernel,
\begin{align}
    K(x,x') = \exp \frac{-\lvert x-x'\rvert}{\lambda},
\end{align}
with scale parameter $\lambda$, and the Inverse Multiquadratic (IMQ) kernel,
\begin{align}
   K(x,x') = \frac{1}{(c + \lvert x-x'\rvert^2)^d},
\end{align}
with parameters $c$ and $d$.

By following common strategies for selecting the parameters, strong performance can be achieved with these kernels without explicit tuning. The median heuristic (taking $\Lambda$ or $\lambda$ equal to the median pairwise distance in the dataset) is a common choice for the Gaussian and exponential kernels. In our case, we actually find that the choice from the main text $\Lambda=1$ performs better and is close to optimal. It corresponds to taking the feature variance as the bandwidth because we preprocess the data such that the features have unit variance.

Table~\ref{tab:chi2_sub} below quantifies the performance of each kernel compared to the AutoDiff loss and OmniFold in terms of the $\chi^2$ metric,
\begin{align}
    \chi^2 = \sum_{i=1}^{N_\text{bins}} \frac{1}{\sigma_\text{data,i}^2}(y_\text{data,i}-y_\text{unfold,i})^2,
\end{align}
with $y_\text{data,i}$ and $y_\text{unfold,i}$ being the histogram values in bin $i$ and $\sigma_\text{data,i}$ the Poisson error on the data. We use $\Lambda,\lambda=1$ such that the Gaussian column corresponds to the results in the main text. For the IMQ kernel, we perform a short scan across $c\in\{0.1,0.5,1.0\}, d\in\{0.5,0.75,0.9\}$ and identify $(c,d)=(0.1,0.9)$ as the best-performing choice in terms of the downstream $\chi^2$ of the classifier score $\log R_\theta(x)$ histogram. Such a scan can be parallelized, and its cost is amortized --- the same kernel can be reused for repeated unfolding, e.g. when varying nuisance parameters. The tuned IMQ kernel achieved slightly better performance compared to the Gaussian or Exponential kernels, though they all improve on 10 iterations of OmniFold. The AutoDiff loss is weaker than the analytic kernels in the marginals of each feature, but achieves the best result on the classifier score.

In tables~\ref{tab:chi2_yukawa}~and~\ref{tab:chi2_full}, we show the $\chi^2$ values for the $tHj$ and constituent-level $Zj$ results respectively.

\begin{table}[bp!]
    \centering
    \begin{tabular}{llccccc}
\toprule
 & & \multicolumn{4}{c}{AUSSIE} & \multirow{2}{*}{OmniFold (It. 10)} \\
\cmidrule{3-6}
 &  & Gaussian & Exponential & Inv. Multiquad. & AutoDiff & \\
\midrule
\multirow{7}{1.6em}{\rotatebox[origin=c]{90}{Reco-level}}
 & $m$                & 1.53                                  & 1.36                                  & \cellcolor{heatgreen!35}\textbf{1.33} & 1.90 & 1.75 \\
 & $N$                & 1.78                                  & \cellcolor{heatgreen!35}\textbf{1.74} & 1.77 & 2.00 & 1.84 \\
 & $w$                & 2.40                                  & 2.52                                  & \cellcolor{heatgreen!35}\textbf{2.25} & 2.50 & 2.38 \\
 & $\tau_{21}$        & 2.24                                  & 2.02                                  & \cellcolor{heatgreen!35}\textbf{1.79} & 2.95 & 5.20 \\
 & $\log \rho$        & 1.44                                  & 1.47                                  & \cellcolor{heatgreen!35}\textbf{1.14} & 2.35 & 1.90 \\
 & $z_g$              & \cellcolor{heatgreen!35}\textbf{1.14} & 1.18                                  & 1.23 & 1.37 & 3.10 \\
 & $\log R_\theta(x)$ & 4.85                                  & 5.87                                  & 3.61 & \cellcolor{heatgreen!35}\textbf{3.18} & 6.34 \\
\midrule
\multirow{6}{1.6em}{\rotatebox[origin=c]{90}{Part-level}}
 & $m$         & \cellcolor{heatgreen!35}\textbf{5.34} & 6.02                                  & 6.25 & 14.44 & 22.93 \\
 & $N$         & 5.36                                  & \cellcolor{heatgreen!35}\textbf{1.33} & 1.79 & 2.36 & 1.94 \\
 & $w$         & \cellcolor{heatgreen!35}\textbf{2.61} & 3.30                                  & 3.21 & 5.68 & 3.19 \\
 & $\tau_{21}$ & 6.53                                  & 3.73                                  & \cellcolor{heatgreen!35}\textbf{3.38} & 12.33 & 39.11 \\
 & $\log \rho$ & 9.40                                  & 10.46                                 & \cellcolor{heatgreen!35}\textbf{8.67} & 11.24 & 10.28 \\
 & $z_g$       & 1.66                                  & 1.87                                  & \cellcolor{heatgreen!35}\textbf{1.17} & 3.45 & 14.41 \\
\bottomrule
\end{tabular}

    \caption{$\chi^2/N_\text{bins}$ values for $Zj$ jet substructure observables. The first column corresponds to the results from Section~\ref{sec:res_sub} of the main text. The best result per row is highlighted.}
    \label{tab:chi2_sub}
\end{table}

\begin{table}[tp]
    \centering
    \begin{tabular}{llcccc}
\toprule
 &  & AUSSIE (AutoDiff) & OmniFold (It. 10)\\
\midrule
\multirow{4}{1.6em}{\rotatebox[origin=c]{90}{Reco-level}} & $E_{j_b}$ & 0.53 & \cellcolor{heatgreen!35}\textbf{0.50} \\
 & $\Delta \eta_{\gamma_2,j_b}$ & \cellcolor{heatgreen!35}\textbf{1.21} & 1.33 \\
 & $\Delta R_{j_1,j_2}$ & \cellcolor{heatgreen!35}\textbf{0.77} & 0.81 \\
 & $\log R_\theta(x)$ & \cellcolor{heatgreen!35}\textbf{1.70} & 6.97 \\
\midrule
\multirow{5}{1.6em}{\rotatebox[origin=c]{90}{Part-level}} & $\eta_{j}$& \cellcolor{heatgreen!35}\textbf{0.96} & 2.12 \\
 & $\Delta R_{t,h}$ & \cellcolor{heatgreen!35}\textbf{0.94} & 1.18 \\
 & $\Delta \eta_{h,j}$ & \cellcolor{heatgreen!35}\textbf{2.25} & 11.77 \\
 & $\Delta \phi_{t,j}$ & \cellcolor{heatgreen!35}\textbf{1.62} & 8.39 \\
 & $\Delta R_{t,j}$ & \cellcolor{heatgreen!35}\textbf{2.89} & 6.68 \\
\bottomrule
\end{tabular}

    \caption{$\chi^2/N_\text{bins}$ values for the $tHj$ dataset in Section~\ref{sec:res_mem}, comparing AUSSIE and various iterations of OmniFold. The best result per row is highlighted.}
    \label{tab:chi2_yukawa}
\end{table}

\begin{table}[bp]
    \centering
    \begin{tabular}{llcccc}
\toprule
 &  & AUSSIE & OmniFold (1) & OmniFold (5) & OmniFold (10)\\
\midrule
\multirow{8}{1.6em}{\rotatebox[origin=c]{90}{Reco-level}} & $p_T$ & \cellcolor{heatgreen!35}\textbf{3.74} & 5.07 & 6.65 & 4.25\\
 & $m$ & 7.43 & 5.25 & \cellcolor{heatgreen!35}\textbf{3.06} & 3.62\\
 & $N$ & 10.07 & 5.05 & \cellcolor{heatgreen!35}\textbf{2.91} & 2.97\\
 & $w$ & 6.83 & 9.60 & 5.70 & \cellcolor{heatgreen!35}\textbf{4.46}\\
 & $\tau_{21}$ & 5.09 & 11.62 & 4.13 & \cellcolor{heatgreen!35}\textbf{3.83}\\
 & $z_g$ & \cellcolor{heatgreen!35}\textbf{1.92} & 6.49 & 3.20 & 2.92\\
 & $\log \rho$ & 8.06 & 10.77 & 8.07 & \cellcolor{heatgreen!35}\textbf{7.03}\\
 & $\log R_\theta(x)$ & \cellcolor{heatgreen!35}\textbf{171.0} & 934.2 & 337.2 & 298.9\\
\midrule
\multirow{7}{1.6em}{\rotatebox[origin=c]{90}{Part-level}} & $p_T$ & 15.15 & \cellcolor{heatgreen!35}\textbf{12.76} & 22.86 & 13.71\\
 & $m$ & 24.63 & 46.32 & 22.46 & \cellcolor{heatgreen!35}\textbf{18.43}\\
 & $N$ & 20.29 & 24.98 & 16.50 & \cellcolor{heatgreen!35}\textbf{16.03}\\
 & $w$ & 6.80 & 12.74 & 5.34 & \cellcolor{heatgreen!35}\textbf{4.28}\\
 & $\tau_{21}$ & \cellcolor{heatgreen!35}\textbf{3.75} & 54.88 & 18.31 & 18.38\\
 & $z_g$ & \cellcolor{heatgreen!35}\textbf{4.68} & 26.39 & 13.52 & 12.60\\
 & $\log \rho$ & \cellcolor{heatgreen!35}\textbf{20.41} & 35.69 & 26.82 & 27.87\\
\bottomrule
\end{tabular}

    \caption{$\chi^2/N_\text{bins}$ values for constituent-level jet substructure unfolding from Section~\ref{sec:res_full}. The best result per row is highlighted.}
    \label{tab:chi2_full}
\end{table}

\clearpage
\bibliography{tilman,literature}

\end{document}